\documentclass[appendixfloats]{emulateapj}
\usepackage[titletoc,title]{appendix}
\usepackage{longtable} 
\usepackage{ltxtable} 
\usepackage{graphicx}
\usepackage{graphics}
\usepackage{color}
\usepackage{epsfig}
\usepackage{rotating}
\usepackage{subfigure}
\usepackage{float}
\usepackage{url}
\usepackage{hyperref}
\usepackage{breakurl}
\usepackage{scalerel}

\shortauthors{J. T. Li et al.}
\shorttitle{CGM-MASS II: Statistical Analaysis of the Sample Galaxies}

\begin{document}

\title{The Circum-Galactic Medium of MASsive Spirals II: Probing the Nature of Hot Gaseous Halo around the Most Massive Isolated Spiral Galaxies}

\author{Jiang-Tao Li\altaffilmark{1}, Joel N. Bregman\altaffilmark{1}, Q. Daniel Wang\altaffilmark{2}, Robert A. Crain\altaffilmark{3}, Michael E. Anderson\altaffilmark{4}, and Shangjia Zhang\altaffilmark{1}} 

\altaffiltext{1}{Department of Astronomy, University of Michigan, 311 West Hall, 1085 S. University Ave, Ann Arbor, MI, 48109-1107, U.S.A.}

\altaffiltext{2}{Department of Astronomy, University of Massachusetts, 710 North Pleasant St., Amherst, MA, 01003, U.S.A.}

\altaffiltext{3}{Astrophysics Research Institute, Liverpool John Moores University, IC2, Liverpool Science Park, 146 Brownlow Hill, Liverpool, L3 5RF, United Kingdom}

\altaffiltext{4}{Max-Planck Institute for Astrophysics, Karl-Schwarzschild-Stra$\rm\beta$e 1, 85748 Garching bei M\"{u}nchen, Germany}


\nonumber

\begin{abstract}
We present the analysis of the \emph{XMM-Newton} data of the {\color{red}C}ircum-{\color{red}G}alactic {\color{red}M}edium of {\color{red}MAS}sive {\color{red}S}pirals (CGM-MASS) sample of six extremely massive spiral galaxies in the local Universe. All the CGM-MASS galaxies have diffuse X-ray emission from hot gas detected above the background extending $\sim(30-100)\rm~kpc$ from the galactic center. This doubles the existing detection of such extended hot CGM around massive spiral galaxies. The radial soft X-ray intensity profile of hot gas can be fitted with a $\beta$-function with the slope typically in the range of $\beta=0.35-0.55$. This range, as well as those $\beta$ values measured for other massive spiral galaxies, including the Milky Way (MW), are in general consistent with X-ray luminous elliptical galaxies of similar hot gas luminosity and temperature, and with those predicted from a hydrostatic isothermal gaseous halo. Hot gas in such massive spiral galaxy tends to have temperature comparable to its virial value, indicating the importance of gravitational heating. This is in contrast to lower mass galaxies where hot gas temperature tends to be systematically higher than the virial one. The ratio of the radiative cooling to free fall timescales of hot gas is much larger than the critical value of $\sim10$ throughout the entire halos of all the CGM-MASS galaxies, indicating the inefficiency of gas cooling and precipitation in the CGM. The hot CGM in these massive spiral galaxies is thus most likely in a hydrostatic state, with the feedback material mixed with the CGM, instead of escaping out of the halo or falling back to the disk. We also homogenize and compare the halo X-ray luminosity measured for the CGM-MASS galaxies and other galaxy samples and discuss the ``missing'' galactic feedback detected in these massive spiral galaxies.
\end{abstract}

\section{Introduction}\label{sec:Introduction}

Isolated spiral galaxies are expected to host hot gaseous halos which can be produced either by various types of galactic feedback or by the accretion and gravitational compression of external gas. Feedback from AGN, supernovae (SNe), or massive stellar winds can produce strong X-ray emission in the halos of galaxies with a broad range of mass (e.g., \citealt{Strickland04,Tullmann06,Li13a}). On the other hand, external gas accreted onto the galaxies can only be heated gravitationally to the virial temperature of the dark matter halo in massive galaxies (via hot mode accretion, e.g., \citealt{Keres09}). Since the radiative cooling curve of typical circum-galactic medium (CGM) peaks at $kT\sim10^{5-6}\rm~K$ where far-UV lines of highly ionized ions emit efficiently (e.g., \citealt{Sutherland93}), only gas at X-ray emitting temperatures above this peak of the cooling curve are expected to be stable in the halo. Therefore, only in a galaxy with mass comparable to or greater than that of the Milky Way (MW) Galaxy (with a rotational velocity of $\sim220\rm~km~s^{-1}$ and a virial temperature of $kT\sim10^{6.3}\rm~K$) do we expect to find a hydrostatic, volume-filling, X-ray-emitting gaseous halo. 

In addition to the instability of the gravitationally heated gas in low- or intermediate-mass halos, another problem preventing us from finding the accreted hot gas is the contamination from feedback material. Archival X-ray observations are often biased to galaxies with high star formation rates (SFRs); only a few observations were available for quiescent ones. These actively star forming galaxies eject chemically enriched gas into their halos, which dominates the X-ray emission around galactic disks (typically within 10-20~kpc). In this case, the accreted gas, although significant in the mass budget, can only radiate in X-ray efficiently after they well mix with the metal enriched feedback material (e.g., \citealt{Crain13}). Therefore, in order to study the effect of gravitational heating of the diffuse X-ray emitting halo gas, we prefer galaxies with low SFR.

Extended X-ray emitting halos have been detected around various types of galaxies (see a review in \citealt{Wang10}). The X-ray luminosity of the halo gas is typically linearly dependent on the disk SFR and is thought to be mostly produced by galactic SNe feedback (e.g., \citealt{Strickland04,Tullmann06,Li08,Li13b,Wang16}), although sometimes Type~Ia SNe from quiescent galaxies may play an important role (e.g., \citealt{Li09,Li15}). Comparison with numerical simulations indicates that models could in general reproduce the X-ray luminosity of $L^\star$ galaxies (e.g., \citealt{Crain10,Li14}). 

On the other hand, the picture is much less clear for spiral galaxies significantly more massive than the MW. Although the hot CGM produced by gravitationally heated externally accreted gas has been predicted many years ago (e.g., \citealt{Benson00,Toft02}), there are just a few deep X-ray observations of massive enough spiral galaxies whose virial temperature is in the X-ray range (e.g., \citealt{Li06,Li07,Rasmussen09,Anderson11,Anderson16,Dai12,Bogdan13,Bogdan15}) and some of them do not have an extended X-ray emitting halo detected significantly beyond the galactic disk and bulge.

We have conducted deep \emph{XMM-Newton} observations of a sample of five (six by adding the archival observation of UGC~12591) massive isolated spiral galaxies in the local Universe [The Circum-Galactic Medium of MASsive Spirals (CGM-MASS) project]. All these galaxies have low SFRs compared to their large stellar masses (Table~\ref{table:GalaxyPara}). An introduction of the sample selection criteria and detailed data reduction procedures, as well as an initial case study of NGC~5908, are presented in \citet{Li16b} (Paper~I). Particularly interesting is that the $L_{\rm X}/M_*$ ratio of this massive isolated spiral galaxy is not significantly higher than those of lower mass non-starburst galaxies. 

Here we present results from the analysis of the \emph{XMM-Newton} data of the whole CGM-MASS sample, including the archival data of UGC~12591 \citep{Dai12}. The reanalysis of this archival data is to make sure that the data reduction and analysis processes are uniform for all the galaxies, which is a key for statistical analysis. 
The paper is organized as follows: In \S\ref{section:Results}, we present the results from analyzing the \emph{XMM-Newton} data of the sample galaxies, including both the spatial and spectral analysis and the derivation of other physical parameters of the hot gas. Some additional details of data analysis, as well as the properties of the prominent extended and point-like X-ray sources in the \emph{XMM-Newton} field of view (FOV), are presented in Appendix~\ref{Appsection:DataReduction}. We then introduce other galaxy samples used for comparison in \S\ref{sec:OtherSample}. We perform statistical analysis comparing the CGM-MASS galaxies to other samples in \S\ref{sec:Statistical} and discuss the scientific implications of the results in \S\ref{section:discussion}. Our main results and conclusions are summarized in \S\ref{section:Summary}. Spatial and spectral analysis based on the stacked data of the whole sample and discussions on the baryon budget will be presented in \citet{Li17} (Paper~III). Errors are quoted at 1~$\sigma$ confidence level throughout the paper unless specifically noted.


\section{Data Analysis of the CGM-MASS Galaxies}\label{section:Results}

\subsection{Multi-wavelength galaxy properties}\label{subsubsec:OtherProperty}

We first update a few parameters of our sample galaxies (Table~\ref{table:GalaxyPara}). In Paper~I, the stellar mass of each galaxies ($M_*$) is estimated from its total K-band magnitude listed in the 2MASS extended source catalog \citep{Skrutskie06}. This magnitude includes the contribution from the galactic nucleus. In the present paper, we exclude the nuclear point-like source and fit the remaining intensity profile along the major axis of the galaxy with an exponential model. The integrated K-band luminosity of this exponential model is then converted to the stellar mass using the same method as adopted in Paper~I. We consider the best estimate as the integration extrapolated into the center. We assume the stellar mass estimated without excluding the nuclear source as the upper limit and the integration without extrapolating on to the center as the lower limit of the estimate. All the stellar mass and its upper and lower limits are calculated within an elliptical region for which the semi-major and semi-minor axis are at the isophotal level of $23\rm~mag~arcsec^{-2}$. 

In Paper~I, the SFR of a galaxy is estimated from its \emph{IRAS} total IR luminosity. We herein update this estimate based on the spatially resolved \emph{WISE} W4 (22~$\rm \mu m$) image, using a similar method as adopted in \citet{Wang16}. 

We caution that the estimate of the stellar mass and SFR may be affected by some systematical biases caused by the enhanced extinction in the edge-on case. Although the CGM-MASS galaxies have low SFRs and cold gas contents so a relatively low extinction especially in IR (Li et al. 2018, in prep), the extinction may not be negligible even in the \emph{WISE} W4 band, as discussed in \citet{Li16a}. Therefore, we do not adopt stellar mass and SFR measurements in shorter wavelength (e.g., \citealt{Maraston13}; Vargas et al. 2017, submitted), which are in general more reliable in face-on cases.

We also obtain the total baryonic mass of the galaxy ($M_{\rm TF}$) from the rotation velocity ($v_{\rm rot}$; Table~1 of Paper~I) using the baryonic Tully-Fisher relation \citep{Bell01}. $M_{\rm TF}$ is a measurement of the gravitational mass, similar as the dark matter halo mass, and is listed here for the ease of comparison with other samples (e.g., \citealt{Li13a}; \S\ref{subsec:EnergyBudget}).

\begin{table}
\vspace{-0.in}
\begin{center}
\caption{Properties of the CGM-MASS Galaxies.} 
\footnotesize
\vspace{-0.0in}
\tabcolsep=1.2pt%
\begin{tabular}{lcccccccccccccc}
\hline\hline
Galaxy        & Scale & $M_*$ & $M_*/L_{\rm K}$ & SFR & $M_{\rm TF}$ \\
            & kpc/arcm & $10^{11}\rm~M_\odot$ & $\rm M_\odot/L_\odot$ & $\rm M_\odot~yr^{-1}$ & $10^{11}\rm~M_\odot$ \\
\hline
UGC 12591 & 27.45 & $5.92_{-0.74}^{+0.14}$ & 0.773 & $1.17\pm0.13$ & $16.1\pm1.5$ \\
NGC 669     & 22.63 & $3.32_{-0.17}^{+0.02}$ & 0.893 & $0.77\pm0.07$ & 5.32 \\
ESO142-G019 & 18.78 & $2.49_{-0.24}^{+0.05}$ & 1.137 & $0.37\pm0.06$ & $5.07\pm0.90$ \\
NGC 5908   & 15.10 & $2.56_{-0.15}^{+0.02}$ & 0.842 & $3.81\pm0.09$ & $4.88\pm0.60$ \\
UGCA 145   & 20.17 & $1.47_{-0.08}^{+0.01}$ & 0.595 & $2.75\pm0.11$ & 4.03 \\
NGC 550     & 27.09 & $2.58_{-0.28}^{+0.04}$ & 0.773 & $0.38\pm0.09$ & $5.08\pm1.81$ \\
\hline\hline
\end{tabular}\label{table:GalaxyPara}
\end{center}
Updated parameters from Paper~I: the stellar mass, $M_*$, measured from the \emph{2MASS} K-band luminosity and the K-band mass-to-light ratio ($M_*/L_{\rm K}$) of the galaxies; $M_*/L_{\rm K}$ is estimated from the inclination, redshift, and Galactic extinction corrected B-V color, except for UGCA~145, for which the corrected B-R color is used (\S\ref{subsubsec:OtherProperty}); SFR estimated from the \emph{WISE} 22~$\rm \mu m$ luminosity (\S\ref{subsubsec:OtherProperty}); the total baryon mass, $M_{\rm TF}$, estimated from the inclination corrected rotation velocity $v_{\rm rot}$ and the baryonic Tully-Fisher relation \citep{Bell01}, and is used to produced Fig.~\ref{fig:Ebudget}b. Some other parameters of the sample galaxies, such as the distance (94.4~Mpc for UGC~12591), $v_{\rm rot}$ ($488.38\pm12.54\rm~km~s^{-1}$ for UGC~12591), $M_{\rm 200}$ ($2.42\times10^{13}\rm~M_\odot$ for UGC~12591), and $r_{\rm 200}$ (601~kpc for UGC~12591), are listed in Paper~I.\\
\end{table}

\subsection{Spatial analysis of the diffuse X-ray emission}\label{subsection:SpatialCorona}

\begin{figure*}
\begin{center}
\epsfig{figure=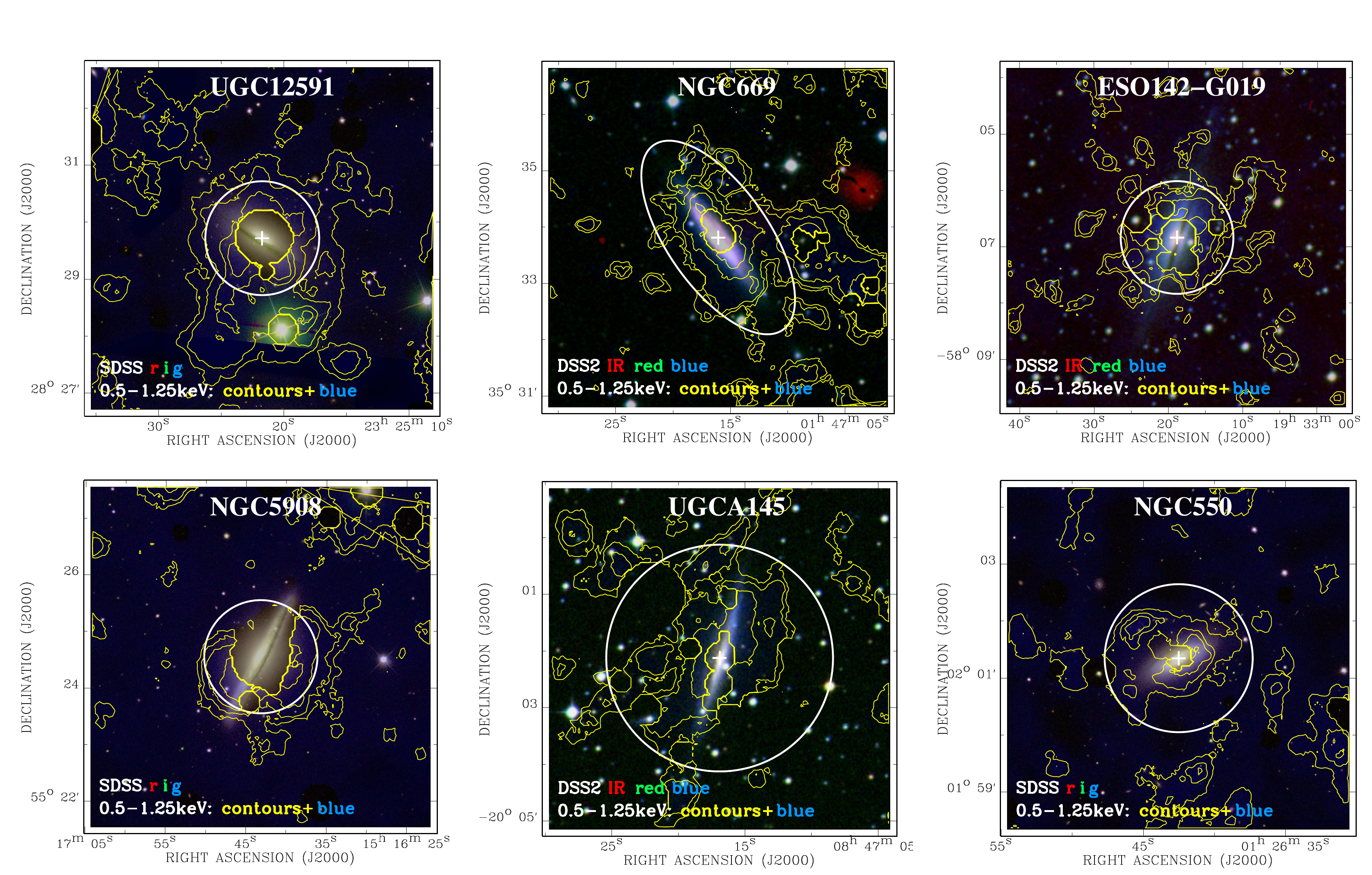,width=1.0\textwidth,angle=0, clip=}
\caption{X-ray contours overlaid on the DSS or SDSS optical tri-color images of the central $6^\prime\times6^\prime$ of the sample galaxies. Contours are the diffuse soft X-ray images at different rms noise levels above the background: 5, 10, 20, 30~$\sigma$ for UGC~12591; 3, 5, 10, 20~$\sigma$ for NGC~669 and ESO142-G019; 2, 3, 5, 10~$\sigma$ for NGC~5908; 3, 5, 10~$\sigma$ for UGCA~145; 10, 15, 20, 30~$\sigma$ for NGC~550. We adopt relatively high $\sigma$ value for NGC~550 because we have removed the bright background cluster Abell~189 (\S\ref{subsection:XMMImages}) when calculating background rms. The white circle or ellipse overlaid in each panel is used to extract the spectra of diffuse X-ray emission from the halo (Fig.~\ref{fig:HaloSpec}).
}\label{fig:imagesZoomin}
\end{center}
\end{figure*}

We present additional information on the \emph{XMM-Newton} data reduction and the results on the prominent extended or point-like sources in Appendix~\ref{Appsection:DataReduction}. We present the major results on the diffuse hot gas emission in the following sections. In Fig.~\ref{fig:imagesZoomin}, we present the point source removed, soft proton and quiescent particle background (QPB) subtracted, exposure corrected, and adaptively smoothed 0.5-1.25~keV \emph{XMM-Newton} image in the central $6^\prime\times6^\prime$ of the CGM-MASS galaxies, in order to show how the diffuse X-ray emission may be associated with the target galaxies.

\begin{figure*}
\begin{center}
\epsfig{figure=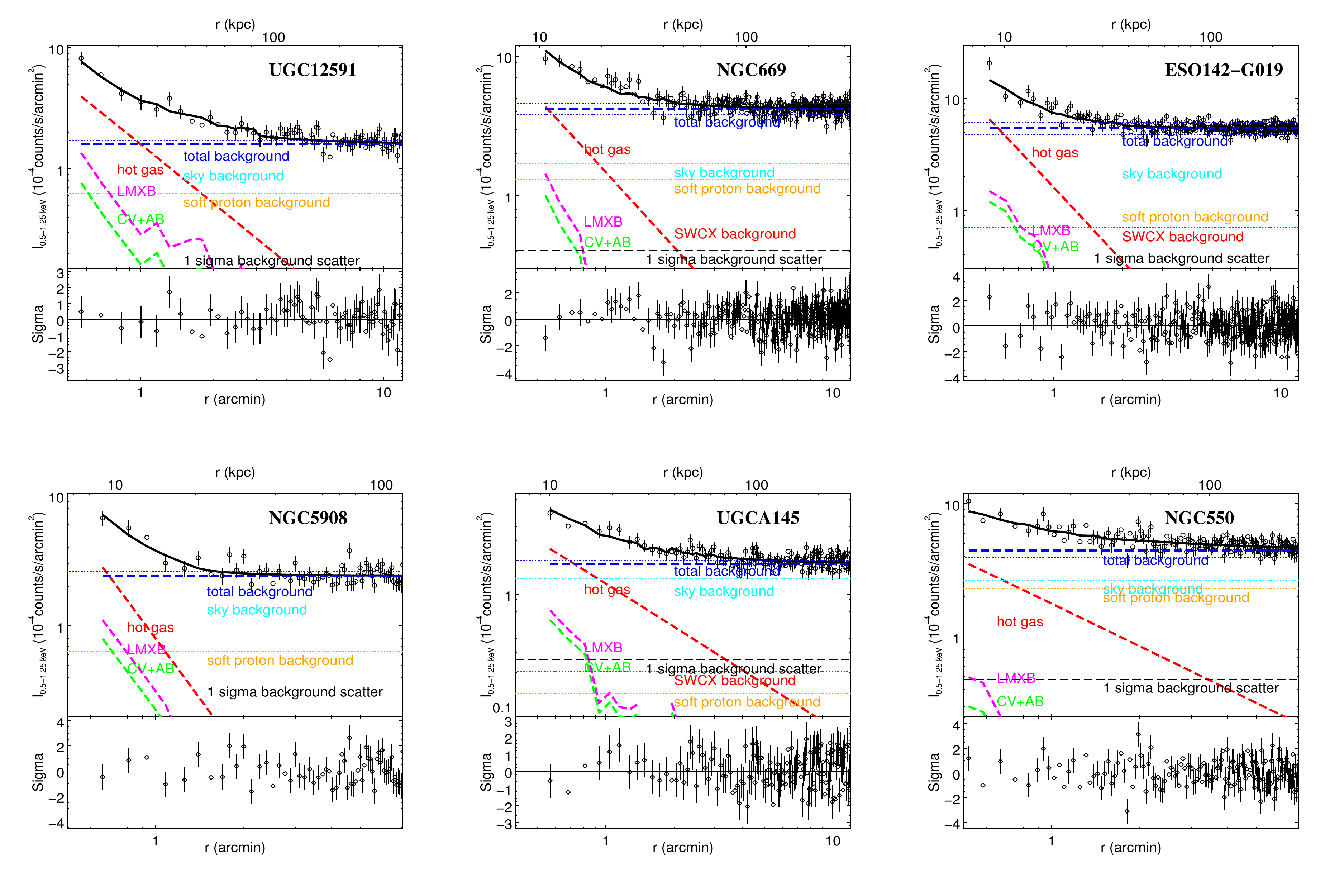,width=1.0\textwidth,angle=0, clip=}
\caption{Radial intensity profile of the diffuse 0.5-1.25~keV emission around the center of the CGM-MASS galaxies. The upper and lower axes denote the off-center distances in unit of kpc and arcmin, respectively. X-ray bright nuclei of the galaxies are masked off (Fig.~\ref{fig:imagesZoomin}), so the profiles typically start at $r\approx0.5^\prime$. The best-fit model, shown as a thick black solid curve, is comprised of several components: the sky+SP+SWCX background (blue dashed), the LMXB and CV+AB contributions estimated from the K-band intensity profile (magenta and green dashed), and a $\beta$-function representing the hot gas distribution (red dashed). Statistical plus systematical uncertainties of the best-fit background level are plotted as blue thin dotted lines. For comparison, we also plot the sky (cyan), SP (orange), and SWCX (red; whenever applicable) background components with thin dotted lines, separately. The 1~$\sigma$ scatter of the background estimated in radial ranges with flat background is also plotted as a thin dashed black line.}\label{fig:RadialProfile}
\end{center}
\end{figure*}

We present QPB-subtracted, exposure corrected 0.5-1.25~keV radial intensity profiles around the centers of the target galaxies in Fig.~\ref{fig:RadialProfile}. X-ray emission in this band has the largest contribution from hot gas and is not seriously affected by the strong instrumental lines (especially the strong Al-K and Si-K lines; Fig.~\ref{fig:bckspec}). The profiles are extracted from the unsmoothed images. We have removed all the detected X-ray point sources and extended X-ray emissions not associated with the target galaxies when creating these radial intensity profiles. Prominent removed extended and point-like features are described in Appendix~\ref{subsection:XMMImages} and \ref{subsection:Xpointsrc}, and the masks used to remove them are presented in Fig.~\ref{fig:SourceDiffuseMask}. The intensity profiles are also regrouped to a signal-to-noise ratio of $\rm S/N>7$ for each bin, where the noise includes the contributions from the removed QPB.

We fit the soft X-ray intensity profile with a $\beta$-function plus various stellar and background components. For all the CGM-MASS galaxies, the deep \emph{XMM-Newton} observations reach a 0.3-7.2~keV point source detection limit of $\sim(2-3)\times10^{38}\rm~ergs~s^{-1}$ (UGC~12591 has a higher value of $\approx5.9\times10^{38}\rm~ergs~s^{-1}$; Table~\ref{table:IdentifiedSources}), allowing us to remove the brightest X-ray sources from the diffuse emission. Below this detection limit, there are still contributions from individually X-ray faint stellar sources, including Low Mass X-ray Binaries (LMXBs) and Cataclysmic Variables plus coronal Active Binaries (CVs+ABs). We scale both the LMXB and CV+AB components to the near-IR (K-band) intensity profile tracing the radiation of old stellar population, using the calibrated ratios from \citet{Gilfanov04} and \citet{Revnivtsev08} and a similar procedure adopted in the study of some quiescent early-type disk galaxies \citet{Li09,Li11}. No contributions from young stellar sources are considered in this paper, which is typically less important at large radii for these quiescent galaxies (see also discussions in \S\ref{subsection:SpecCorona}). 

After subtracting the QPB, the residual X-ray background typically includes two components: the cosmic X-ray background produced by the local hot bubble, the Milky Way halo, and distant AGN (e.g., \citealt{Li08}), and the residual soft proton contribution (e.g., \citealt{Kuntz08}). For some galaxies, we also add a background component from SWCX. Detailed background analysis of the sample galaxies are presented in \S\ref{subsection:SWCX} and Fig.~\ref{fig:bckspec}. In analysis of the radial intensity profile, we directly fit the background with a constant level. This best-fit background level is in general consistent with expected from the summation of different background components (sky, soft proton, and sometimes SWCX). However, there may be some systematical uncertainties of the background, such as the intrinsic uncertainties of different stellar and background components, which are difficult to quantify. We roughly characterize this systematical uncertainty using the standard deviation of the total background level estimated in three different ways: (1) the direct fit with the stellar components fixed as presented in Fig.~\ref{fig:RadialProfile}; (2) fit with the stellar components allowed to vary for 50\%; (3) the summation of the rescaled sky, soft proton, and SWCX background components from spectral analysis (also marked in Fig.~\ref{fig:RadialProfile}). The systematical uncertainty estimated this way is typically comparable to or larger than the 1~$\sigma$ statistical error. The total systematical and statistical uncertainties of the background are plotted in Fig.~\ref{fig:RadialProfile}, in comparison with the 1~$\sigma$ background fluctuation.

The best-fit models of the radial intensity profiles are presented in Fig.~\ref{fig:RadialProfile}. The hot gas component is fitted with a $\beta$-function:
\begin{equation}\label{equi:Ibeta}
I=I_{\rm 0}[1+(r/r_{\rm c})^2]^{0.5-3\beta},
\end{equation}
where $I_{\rm 0}$ is the X-ray intensity at $r=0$. As shown in Fig.~\ref{fig:RadialProfile}, due to the presence of X-ray bright sources in the nuclear region, the radial intensity profiles are extracted typically at $r\gtrsim0.5^\prime$. Therefore, the core radius $r_{\rm c}$ of the $\beta$-function is poorly constrained and only affect $I_{\rm 0}$ (not $\beta$) of the $\beta$-function. We then fix $r_{\rm c}$ at $0.1^\prime$ which is much smaller than the radius of the removed nuclear region of the AGN. The best fit values of $I_{\rm 0}$ (depends on the assumed $r_{\rm c}$) and $\beta$ are listed in Table~\ref{table:SpecPara}. Extended diffuse soft X-ray emission can typically be detected above the 1~$\sigma$ scatter of the background to $r\sim(30-100)\rm~kpc$ around individual galaxies (Fig.~\ref{fig:RadialProfile}). The slope of the radial intensity profile is typically $\beta\sim0.5$, with NGC~5908 studied in Paper~I has the steepest radial intensity distribution ($\beta\approx0.68$). For these quiescent galaxies, there is no significant evidence of the variation of the slope of the radial intensity profile at $r\approx(10-100)\rm~kpc$. We will discuss the radial distribution of hot gas based on the stacked X-ray intensity profile in Paper~III. 

\subsection{Spectral analysis of the diffuse X-ray emission}\label{subsection:SpecCorona}

We extract diffuse X-ray spectra of individual CGM-MASS galaxies after subtracting the detected X-ray point-like sources and the unrelated prominent diffuse X-ray features from the circular or elliptical regions shown in Fig.~\ref{fig:imagesZoomin}. These spectral analysis regions enclose the most prominent diffuse X-ray features associated with the galaxy.

\begin{figure*}
\begin{center}
\epsfig{figure=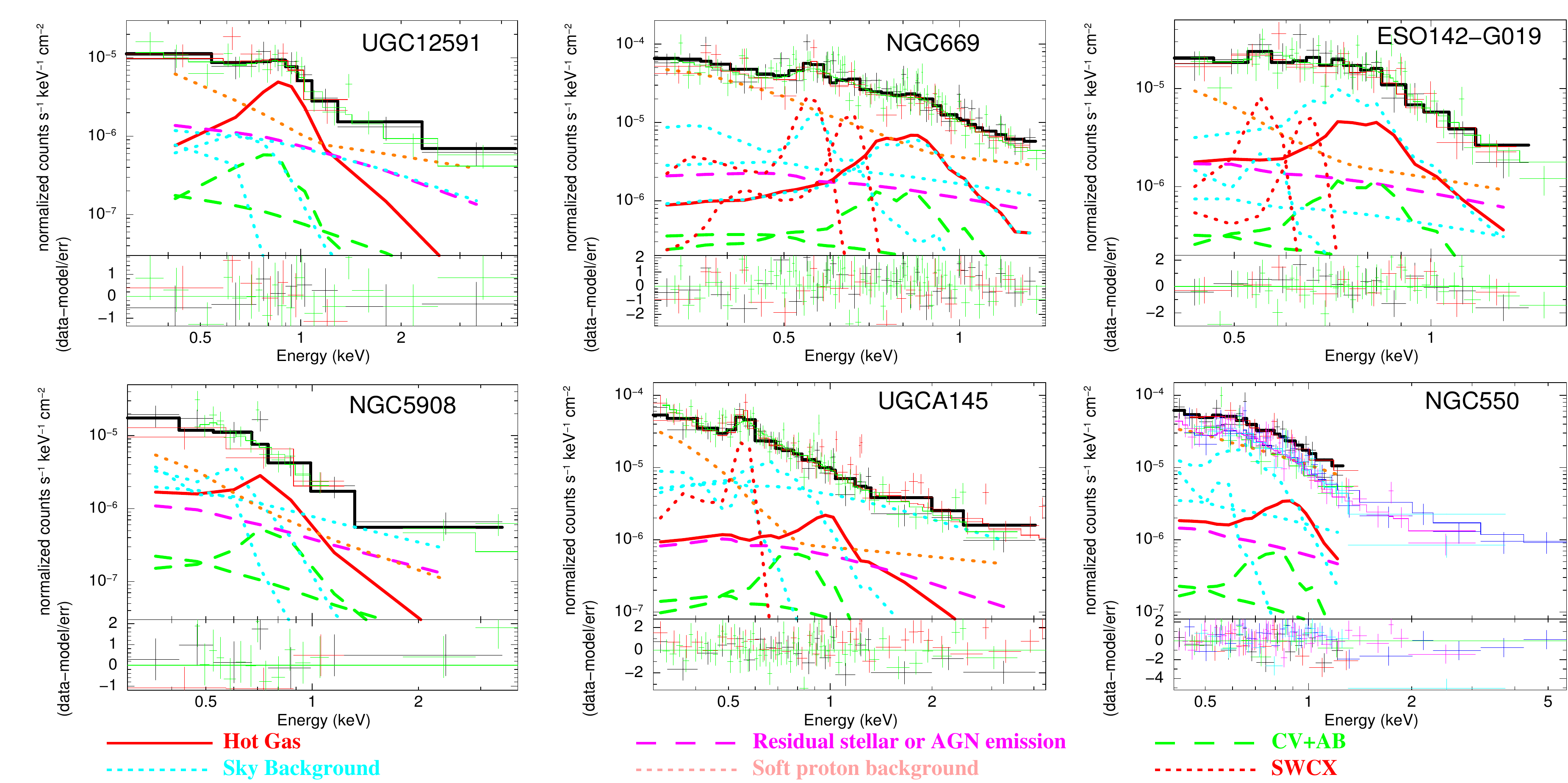,width=1.0\textwidth,angle=0, clip=}
\caption{Diffuse X-ray spectra extracted from the inner halos of our sample galaxies (white circle or ellipse in Fig.~\ref{fig:imagesZoomin}). All the spectra have been regrouped to achieve a $\rm S/N>3$. Curves representing different model components are denoted at the bottom of the figure. The plotted data points have been scaled with the effective area of each instrument (MOS-1: black; MOS-2: red; PN: green), so they are consistent with the summation of various model components. There are two observations of NGC~550. Data from the second observation are plotted in blue (MOS-1), cyan (MOS-2), and magenta (PN), respectively. }\label{fig:HaloSpec}
\end{center}
\end{figure*}

We rescale all the background model components (according to the area scale of the spectral analysis regions), as well as the LMXB and the CV+AB components (according to the K-band luminosity enclosed by the spectral analysis regions), and add them to the model of the source spectra. Model parameters of different background and stellar components are discussed in Appendix~\ref{subsection:SWCX} and Paper~I. In particular, we adopt a $\Gamma=1.6$ power law to model the LMXB component, and a $kT=0.5\rm~keV$ thermal plasma plus a $\Gamma=1.9$ power law to model the CV+AB component. The normalization of these model components are computed with the stellar mass enclosed by the spectral analysis regions. Young stellar contributions, such as high mass X-ray binaries (HMXBs), are difficult to quantify, because they mainly distribute in the galactic disk or nuclear regions, which are largely removed in spectral analysis and highly absorbed in edge-on cases. Using a similar procedure as Paper~I and adopting the new estimates of SFRs (Table~\ref{table:GalaxyPara}), we can compute the upper limits of HMXB contributions (without considering the removed regions or the additional absorption by the cool gas in the galactic disk), which is $\sim10^{38-39}\rm~ergs~s^{-1}$ in 0.5-2~keV for the CGM-MASS galaxies. This luminosity is typically $\lesssim10\%$ of the diffuse hot gas emission (Table~\ref{table:SpecPara}). We therefore do not consider an additional component describing the young stellar contribution in these extremely quiescent galaxies. All parameters of the background and stellar components discussed above are fixed.

We model the hot gas emission with an ``{\small APEC}'' model, which is subjected to absorption at a column density of the Galactic foreground value (listed in Table~1 of Paper~I). The metallicity of hot gas is poorly constrained, so we fix it at $0.2\rm~Z_\odot$, which is consistent with some recent estimates (e.g., \citealt{Bogdan13,Anderson16}). We also add a gain correction to the response file of the PN spectrum (``{\small GAIN}'' model in XSpec), in order to account for the deficiency in the low-energy calibration of the PN camera \citep{Dennerl04}. Such a gain correction has been proved to be important in analyzing the \emph{XMM-Newton} data taken in recent years (e.g., \citealt{Li15b,Li16c}). The slope of the {\small GAIN} is fixed at 1 and the offset is set free. Finally, there are only three free parameters: the temperature and normalization (or X-ray luminosity) of {\small APEC} and the offset of {\small GAIN}. The fitted spectra of each galaxy are presented in Fig.~\ref{fig:HaloSpec} and the best-fit hot gas temperature and 0.5-2.0~keV luminosity ($L_{\rm X, r<r_{spec}}$) are summarized in Table~\ref{table:SpecPara}.

We also analyze the diffuse X-ray spectra extracted from larger radii, but the hot gas emission is too weak compared to various background components (see Fig.~\ref{fig:RadialProfile} for their levels) and the counting statistic is also poor. The measured hot gas properties are largely uncertain. Therefore, in the following discussions, we assume constant hot gas temperature and metallicity, and estimate the X-ray emission of hot gas based on extrapolation of the best-fit radial intensity profile.

In the next few sections, we will statistically compare the X-ray luminosity of the hot halo measured at small and large radii to other samples. However, X-ray emission of the CGM-MASS galaxies are only directly detected to $r=(0.1-0.2)r_{\rm 200}$ ($r_{\rm 200}\sim350-600\rm~kpc$ for the CGM-MASS galaxies; Paper~I). We therefore need to rescale the directly measured hot gas luminosity in the spectral analysis region using the best-fit $\beta$-function of the radial intensity profiles (Equ.~\ref{equi:Ibeta}). By doing such rescaling, we have assumed Equ.~\ref{equi:Ibeta} can be extrapolated to both smaller and larger radii, which further means there is no significant contribution from young stellar sources within the galactic disk and the hot gas within the entire halo follows the same slope of radial distribution as the gas close to the galaxy's stellar content. These assumptions will be further discussed in Paper~III. The rescaled $L_{\rm X}$ at $r<1^\prime$, $r<0.1r_{\rm 200}$, and $r=(0.05-0.15)r_{\rm 200}$ are also listed in Table~\ref{table:SpecPara}.

\subsection{Derived hot gas properties}\label{subsection:DerivedProperties}

We estimate the hot gas properties at a given radius, based on the results from the above spatial (\S\ref{subsection:SpatialCorona}) and spectral analyses (\S\ref{subsection:SpecCorona}), and following a similar procedure as described in \citet{Ge16}. Assuming spherical symmetric distribution of the hot halo gas which also follows a $\beta$-model in radial distribution, the deprojected hydrogen density profile can be written as: 
\begin{equation}\label{equi:nHbeta}
n_{\rm H}=n_{\rm 0}[1+(r/r_{\rm c})^2]^{-\frac{3}{2}\beta},
\end{equation}
where $r_{\rm c}$ and $\beta$ are the same as in Equ.~\ref{equi:Ibeta}, and $n_{\rm 0}$ is the hydrogen number density at $r=0$. Assuming the temperature and metallicity of hot gas do not change with the galactocentric radius, $n_{\rm 0}$ can be expressed with the parameters of the radial intensity profile and the spectral models ($I_{\rm 0}$, $r_{\rm c}$, $\beta$, $\frac{n_{\rm e}}{n_{\rm H}}$, $\frac{CR}{N_{\rm APEC}}$) in the following form (converted from Equ.~10 of \citealt{Ge16}):
\begin{equation}\label{equi:n0}
n_{\rm 0}=0.123\pi^{-3/4}[\frac{I_{\rm 0}}{\frac{CR}{N_{\rm APEC}}\frac{n_{\rm e}}{n_{\rm H}}r_{\rm c}}\frac{\Gamma(3\beta)}{\Gamma(3\beta-0.5)}]^{1/2}
\end{equation}
where $\Gamma$ is the gamma function; $n_{\rm 0}$ is in unit of $\rm cm^{-3}$; $I_{\rm 0}$ in $\rm counts~s^{-1}~arcmin^{-2}$, $r_{\rm c}$ in Mpc, and $\beta$ are obtained from fitting the radial intensity profile; $\frac{n_{\rm e}}{n_{\rm H}}$ is the electron to hydrogen number density ratio at a given metallicity, assuming all the elements are fully ionized; $\frac{CR}{N_{\rm APEC}}$ is the conversion factor from the normalization of the APEC model to the counts rate (0.5-1.25~keV, scaled to MOS-2 with a medium filter) in unit of $\rm counts~s^{-1}~cm^5$. $\frac{n_{\rm e}}{n_{\rm H}}$ and $\frac{CR}{N_{\rm APEC}}$ depend on the spectral model. They are calculated from the best-fit {\small APEC} model describing the hot gas emission of each galaxy and are assumed to be constant at different radii. 

The thermal pressure of hot gas at a given radius can be expressed as:
\begin{equation}\label{equi:Pbeta}
P=n_{\rm t}k_{\rm B}T=P_{\rm 0}[1+(r/r_{\rm c})^2]^{-\frac{3}{2}\beta},
\end{equation}
where $n_{\rm t}=n_{\rm i}+n_{\rm e}$ is the total particle number density including both ions and electrons, $k_{\rm B}$ is the Boltzmann constant and $T$ is the temperature of the hot gas. Similar as $\frac{n_{\rm e}}{n_{\rm H}}$, $\frac{n_{\rm t}}{n_{\rm e}}$ also depends on the spectral model. The thermal pressure at $r=0$ can be expressed with $n_{\rm 0}$ as:
\begin{equation}\label{equi:P0}
P_{\rm 0}=n_{\rm 0}\frac{n_{\rm t}}{n_{\rm e}}\frac{n_{\rm e}}{n_{\rm H}}k_{\rm B}T.
\end{equation}

We also calculate the radiative cooling timescale of the hot gas based on the density profile:
\begin{equation}\label{equi:tcoolbeta}
t_{\rm cool}=\frac{U}{\Lambda_{\rm N}n_{\rm i}n_{\rm e}}=t_{\rm cool,0}[1+(r/r_{\rm c})^2]^{\frac{3}{2}\beta},
\end{equation}
where, $U=\frac{3}{2}n_{\rm t}k_{\rm B}T$ is the internal energy of the hot gas, $\Lambda_{\rm N}$ is the normalized radiative cooling rate in unit of $\rm ergs~s^{-1}~cm^3$. We adopt $\log\Lambda_{\rm N}/{\rm (ergs~s^{-1}~cm^3)}=-23.00$ for the $T=10^{6.65}\rm~K$, $\rm [Fe/H]=-1.0$ thermal plasma from \citet{Sutherland93}. The radiative cooling timescale at $r=0$ can be expressed as:
\begin{equation}\label{equi:tcool0}
t_{\rm cool,0}=\frac{3k_{\rm B}T}{2\Lambda_{\rm N}}\frac{\frac{n_{\rm t}}{n_{\rm e}}}{\frac{n_{\rm e}}{n_{\rm H}}(\frac{n_{\rm t}}{n_{\rm e}}-1)}\frac{1}{n_{\rm 0}}.
\end{equation}

Using Equ.~\ref{equi:tcoolbeta}, we can derive the cooling radius $r_{\rm cool}$ which is defined as the galactocentric radius at which the radiative cooling timescale equals to the Hubble time $t_{\rm Hubble}$. $r_{\rm cool}$ can be expressed as:
\begin{equation}\label{equi:rcool}
r_{\rm cool}=r_{\rm c}[(\frac{t_{\rm Hubble}}{t_{\rm cool,0}})^{\frac{2}{3\beta}}-1]^{\frac{1}{2}}.
\end{equation}

We also derive the column density of hot gas at a given projected distance from the galactic center:
\begin{equation}\label{equi:Npbeta}
N_{\rm p}=N_{\rm p,0}[1+(r/r_{\rm c})^2]^{\frac{1}{2}-\frac{3}{2}\beta},
\end{equation}
where $N_{\rm p,0}$ is the column density along the sightline through the galactic center, and can be expressed as \citep{Ge16}:
\begin{equation}\label{equi:Np0}
N_{\rm p,0}/(10^{20}\rm cm^{-2})=5.47\times10^4n_{\rm 0}r_{\rm c}\frac{\Gamma(3\beta/2-1/2)}{\Gamma(3\beta/2)},
\end{equation}
where $n_{\rm 0}$ is in unit of $\rm cm^{-3}$ and $r_{\rm c}$ in Mpc. Equ.~\ref{equi:Np0} is valid for $\beta>1/3$ \citep{Ge16}.

By integrating the density and energy density profiles (differs from the pressure profile by a factor of $\frac{3}{2}$) of the hot gas, we can obtain the total mass and thermal energy of hot gas within a given radius:
\begin{equation}\label{equi:mass}
M_{\rm hot}=4\pi n_{\rm 0}r_{\rm c}^3\int_{0}^{x}(1+x^2)^{-\frac{3}{2}\beta}x^2dx,
\end{equation}
\begin{equation}\label{equi:energy}
E_{\rm hot}=6\pi P_{\rm 0}r_{\rm c}^3\int_{0}^{x}(1+x^2)^{-\frac{3}{2}\beta}x^2dx,
\end{equation}
where $x=r/r_{\rm c}$. The integral part containing the dimensionless parameter $x$ can be computed with numerical integration. 

We can also compute the radiative cooling rate of the hot gas, which is defined as: $\dot{M}_{\rm cool}=M_{\rm hot}/t_{\rm cool}$. $\dot{M}_{\rm cool}$ can be computed with:
\begin{equation}\label{equi:Mdotcool}
\dot{M}_{\rm cool}=\frac{4\pi n_{\rm 0}r_{\rm c}^3}{t_{\rm cool,0}}\int_{0}^{x}(1+x^2)^{-3\beta}x^2dx.
\end{equation}

We list $n_{\rm 0}$, $P_{\rm 0}$, $t_{\rm cool,0}$, $r_{\rm cool}$ (calculated assuming $t_{\rm Hubble}=10\rm~Gyr$), $N_{\rm p,0}$, as well as $M_{\rm hot}$ and $E_{\rm hot}$ calculated at $r\leq r_{\rm 200}$ and $\dot{M}_{\rm cool}$ calculated at $r\leq r_{\rm cool}$ in Table~\ref{table:LXRadii}. We have adopted $\frac{n_{\rm e}}{n_{\rm H}}=1.20$ and $\frac{n_{\rm t}}{n_{\rm e}}=1.92$ which are calculated assuming 0.2~solar metallicity of the hot gas. We do not account for the error of $\beta$ when computing the error of $M_{\rm hot}$, $E_{\rm hot}$, and $\dot{M}_{\rm cool}$ using the integration term. 

\begin{table*}[t]{}
\vspace{-0.in}
\begin{center}
\caption{Parameters of the Hot Gas Component} 
\footnotesize
\vspace{-0.0in}
\tabcolsep=3.5pt%
\begin{tabular}{lcccccccccccc}
\hline \hline
Galaxy        & $I_{\rm 0}$ & $\beta$ & $r_{\rm spec}$ & $kT$ & $L_{\rm X, r<r_{spec}}$ & $L_{\rm X, r<1^\prime}$ & $L_{\rm X, r<0.1r_{\rm 200}}$ & $L_{\rm X, (0.05-0.15)r_{\rm 200}}$ \\
 & $\rm 10^{-2}counts/s/arcmin^{2}$ & & arcmin & keV & $10^{39}\rm ergs~s^{-1}$ & $10^{39}\rm ergs~s^{-1}$ & $10^{39}\rm ergs~s^{-1}$ & $10^{39}\rm ergs~s^{-1}$ \\
\hline
UGC 12591      & $0.70_{-0.33}^{+0.64}$ & $0.44\pm0.05$ & 1.0 & $0.86_{-0.10}^{+0.08}$ & $4.41_{-0.77}^{+0.33}$ & $11.57_{-2.01}^{+0.87}$ & $18.16_{-3.16}^{+1.37}$ & $10.27_{-1.79}^{+0.77}$ \\
NGC 669          & $0.93_{-0.36}^{+0.61}$ & $0.47\pm0.04$ & $2.0\times0.9$ & $0.68_{-0.10}^{+0.08}$ & $3.20_{-0.57}^{+0.39}$ & $5.33_{-0.95}^{+0.65}$ & $6.48_{-1.16}^{+0.79}$ & $3.12_{-0.56}^{+0.38}$ \\
ESO142-G019 & $2.54_{-1.22}^{+2.56}$ & $0.53\pm0.05$ & 1.0 & $0.68_{-0.10}^{+0.06}$ & $1.57_{-0.27}^{+0.30}$ & $4.88_{-0.84}^{+0.93}$ & $6.13_{-1.05}^{+1.17}$ & $1.61_{-0.28}^{+0.31}$ \\
NGC 5908        & $9.67_{-7.32}^{+43.89}$ & $0.68_{-0.11}^{+0.14}$ & 1.0 & $0.38_{-0.09}^{+0.64}$ & $0.46_{-0.15}^{+0.18}$ & $6.83_{-2.20}^{+2.73}$ & $7.24_{-2.33}^{+2.89}$ & $0.31_{-0.10}^{+0.12}$ \\
UGCA 145       & $0.24_{-0.09}^{+0.15}$ & $0.38\pm0.03$ & 2.0 & $1.08_{-0.15}^{+0.18}$ & $1.45_{-0.44}^{+0.42}$ & $2.11_{-0.64}^{+0.61}$ & $2.07_{-0.62}^{+0.60}$ & $1.69_{-0.51}^{+0.49}$ \\
NGC 550         & $0.21_{-0.09}^{+0.17}$ & $0.35\pm0.05$ & 1.3 & $0.86_{-0.12}^{+0.07}$ & $2.87_{-0.60}^{+0.70}$ & $3.38_{-0.70}^{+0.82}$ & $3.65_{-0.76}^{+0.88}$ & $3.70_{-0.77}^{+0.90}$ \\
\hline \hline
\end{tabular}\label{table:SpecPara}
\end{center}
$I_{\rm 0}$ and $\beta$ are parameters of the $\beta$-function used to fit the radial intensity distribution of the hot gas component (Equ.~\ref{equi:Ibeta}), where $r_{\rm c}$ is fixed at $0.1^\prime$. $r_{\rm spec}$ is the outer radius (or the major and minor radius of the elliptical region of NGC~669) of the spectral analysis regions as plotted in Fig.~\ref{fig:imagesZoomin}. $kT$ is the hot gas temperature measured within the spectral analysis region. $L_{\rm X}$ is measured in 0.5-2.0~keV after correcting the Galactic foreground extinction. $L_{\rm X, r<r_{spec}}$ is directly measured within the spectral analysis regions, while $L_{\rm X, r<1^\prime}$, $L_{\rm X, r<0.1r_{\rm 200}}$, and $L_{\rm X, (0.05-0.15)r_{\rm 200}}$ have been rescaled to different galactocentric radii, assuming the best-fit $\beta$-function of the radial intensity profile, after correcting for the removed point sources or extended features. 
\end{table*}

\begin{table*}[t]{}
\vspace{-0.in}
\begin{center}
\caption{Derived Parameters of the Hot Gas Component} 
\footnotesize
\vspace{-0.0in}
\tabcolsep=2.5pt%
\begin{tabular}{lcccccccccccc}
\hline \hline
Galaxy        &  $n_{\rm 0}$ & $P_{\rm 0}$ & $t_{\rm cool, 0}$ & $M_{\rm hot, r<r_{200}}$ & $E_{\rm hot, r<r_{200}}$ &  $r_{\rm cool}$ & $N_{\rm p,0}$ & $\dot{M}_{\rm cool, r<r_{cool}}$ \\
 & $10^{-3}f^{-1/2}\rm cm^{-3}$ & $f^{-1/2}\rm eV~cm^{-3}$ & $f^{1/2}\rm Gyr$ & $10^{11}f^{1/2}\rm M_\odot$ & $10^{59}f^{1/2}\rm erg$ & kpc & $10^{20}f^{-1/2}\rm cm^{-2}$ & $\rm M_\odot~yr^{-1}$ \\
\hline
UGC 12591      & $7.53_{-1.82}^{+3.45}$ & $12.79_{-3.41}^{+5.97}$ & $1.51_{-0.71}^{+0.39}$ & $3.08_{-0.74}^{+1.41}$ & $11.85_{-3.16}^{+5.53}$ & $11.17_{-8.37}^{+15.05}$ & $4.84_{-1.74}^{+3.77}$ & $0.062_{-0.022}^{+0.041}$ \\
NGC 669          & $8.33_{-1.65}^{+2.73}$ & $11.20_{-2.73}^{+3.91}$ & $1.08_{-0.39}^{+0.25}$ & $1.05_{-0.21}^{+0.34}$ & $3.19_{-0.78}^{+1.11}$ & $10.86_{-2.81}^{+3.91}$ & $3.66_{-1.00}^{+1.65}$ & $0.054_{-0.017}^{+0.026}$ \\
ESO142-G019 & $16.73_{-4.09}^{+8.43}$ & $22.63_{-6.41}^{+11.56}$ & $0.54_{-0.28}^{+0.14}$ & $0.61_{-0.15}^{+0.31}$ & $1.88_{-0.53}^{+0.96}$ & $11.49_{-2.24}^{+2.14}$ & $4.43_{-1.31}^{+2.48}$ & $0.10_{-0.04}^{+0.07}$ \\
NGC 5908        & $46.26_{-17.78}^{+105.04}$ & $34.48_{-15.58}^{+97.77}$ & $0.11 (<0.30)$ & $0.14_{-0.06}^{+0.33}$ & $0.24_{-0.11}^{+0.69}$ & $13.85_{-6.43}^{+4.91}$ & $6.60_{-2.87}^{+15.13}$ & $0.37(<1.55)$ \\
UGCA 145        & $4.76_{-0.95}^{+1.47}$ & $10.22_{-2.46}^{+3.59}$ & $3.02_{-1.02}^{+0.79}$ & $1.46_{-0.29}^{+0.45}$ & $7.08_{-1.71}^{+2.49}$ & $5.37(<91.24)$  & $4.49_{-1.94}^{+10.04}$ & $0.006_{-0.002}^{+0.003}$ \\
NGC 550          & $3.04_{-0.74}^{+1.26}$ & $5.15_{-1.44}^{+2.17}$ & $3.75_{-1.64}^{+0.96}$ & $1.98_{-0.48}^{+0.82}$ & $7.59_{-2.13}^{+3.20}$ & $6.45(<532.5)$  & $14.72_{-12.13}^{+19.85}$ & $0.007_{-0.003}^{+0.004}$ \\
\hline \hline
\end{tabular}\label{table:LXRadii}
\end{center}
$n_{\rm 0}$, $P_{\rm 0}$, $t_{\rm cool, 0}$, $N_{\rm p,0}$ are the hydrogen number density, thermal pressure, radiative cooling timescale, and hydrogen column density of the hot gas at the center of the galaxy ($r=0$), which, together with $\beta$ and $r_{\rm c}$, can be used to characterize the radial distribution of hot gas properties (Equ.~\ref{equi:nHbeta}, \ref{equi:Pbeta}, \ref{equi:tcoolbeta}, \ref{equi:Npbeta}). $M_{\rm hot, r<r_{200}}$ and $E_{\rm hot, r<r_{200}}$ are the total mass and thermal energy of the hot gas integrated within $r_{\rm 200}$. $r_{\rm cool}$ is the cooling radius defined as where the radiative cooling timescale equals to 10~Gyr, assuming the volume filling factor of the soft X-ray emitting hot gas $f=1$. $\dot{M}_{\rm cool, r<r_{cool}}$ is the integrated radiative cooling rate calculated within $r_{\rm cool}$.
\end{table*}


\section{Samples Used for Comparison and Data Homogenization}\label{sec:OtherSample}

\subsection{Nearby highly inclined disk galaxies}\label{subsec:Li13Sample}

There are several systematic studies of the hot gas emission in the halo of nearby galaxies (e.g., \citealt{Strickland04,Tullmann06}), but the samples are either small or the characterization of the hot halo properties are not uniform to compare with other galaxies. We herein mainly compare our CGM-MASS galaxies to the \emph{Chandra} sample of 53 nearby highly inclined disk galaxies studied in \citet{Li13a,Li13b,Li14,Wang16}. The soft X-ray luminosity of hot gas in this sample has been rescaled for a uniform comparison with numerical simulations from \citet{Crain10} in \citet{Li14}, so we refer this sample as ``Li14'' hereafter.

\citet{Li13a} fitted the vertical and horizontal soft X-ray intensity profiles of their sample galaxies with exponential functions. We then rescale the halo X-ray luminosity of Li14's sample to $h<5h_{\rm scal}$ in the vertical direction and $r<5r_{\rm scal}$ in the horizontal direction (along the disk), where $h_{\rm scal}$ and $r_{\rm scal}$ represents the scale height in the vertical direction and the scale length in the horizontal direction, respectively. We also estimate $r_{\rm 200}$ and $M_{\rm 200}$ of Li14's sample in the same way as for the CGM-MASS galaxies, but we caution that some of Li14's sample galaxies show structures of tidal interactions, so the rotation velocity may not exactly reflect the depth of the gravitational potential. Therefore, we rescale the X-ray luminosity according to the directly measured $r_{\rm scal}$ instead of $r_{\rm 200}$. The typical value of $5r_{\rm scal}$ is comparable to $0.1r_{\rm 200}$ \citep{Li13a} where we have rescaled the X-ray luminosity of the CGM-MASS galaxies too. Since X-ray emission declines fast toward large radii, slightly change of the outer radius of the rescaling region does not affect the rescaled X-ray luminosity significantly. The comparison of X-ray emission in the inner region of the dark matter halo between Li14's and the CGM-MASS samples is therefore uniform.

Many of Li14's galaxies do not have enough counts to estimate the temperature of hot gas. For those with a temperature estimation from spectral analysis, the X-ray spectrum is typically extracted within a few tens of kpc from the galactic center, comparable to the CGM-MASS galaxies. We therefore use the directly measured hot gas temperature of Li14's sample for statistical comparisons.

\subsection{Other massive spiral galaxies}\label{subsec:MassiveSpiralSample}

There are very few direct detection of the extended X-ray emission around massive isolated spiral galaxies. Some examples include \citet{Anderson11,Anderson16,Dai12,Bogdan13}. We include the two best cases, NGC~1961 \citep{Bogdan13,Anderson16} and NGC~6753 \citep{Bogdan13}, for comparison here. We convert the X-ray luminosity measured by \citet{Bogdan13} at $r<50\rm~kpc$ to $r<0.1r_{\rm 200}$ using their best-fit modified $\beta$-function (different from Equ.~\ref{equi:Ibeta}). As the modified $\beta$-function strongly overpredict the X-ray emission at extremely small radii, we only integrate the X-ray intensity profile between $r=(0.05-0.1)r_{\rm 200}$, where the X-ray intensity profiles are well fitted with the model. For $r<0.05r_{\rm 200}$, we instead adopt a $\beta$-function with $\beta=0.47$ and $r_{\rm core}=1\rm~kpc$ from \citet{Anderson11}. The X-ray luminosity estimated this way is $L_{\rm X, r<0.1r_{\rm 200}}=(7.80\pm2.23)\times10^{40}\rm~ergs~s^{1}$ for NGC~1961 and $(9.38\pm1.51)\times10^{40}\rm~ergs~s^{1}$ for NGC~6753.

For the slope of the radial intensity profile ($\beta$), we adopt the value of NGC~1961 from \citet{Anderson11} ($0.47_{-0.06}^{+0.07}$). For NGC~6753, since the slope of \citet{Bogdan13}'s modified $\beta$-function ($\beta_{\rm modify}$) approaches to $\beta+0.5/3.0$ at large radii (where $\beta$ is the slope of the standard $\beta$-function of Equ.~\ref{equi:Ibeta}), we obtain $\beta=0.54$ from the originally measured $\beta_{\rm modify}$ of 0.37. 

We adopt the temperature of the hot gas measured in $r=(0.05-0.15)r_{\rm 200}$ from \citet{Bogdan13}. This temperature may be slightly biased when compared to the temperatures measured at smaller radii (such as for the CGM-MASS galaxies), but we do not find any significant evidence of radial variation of the hot gas temperature for these massive spiral galaxies. 

X-ray emission detected at larger radii may have different properties from those detected close to the galaxy's stellar content. \citet{Bogdan15} present upper limits of the X-ray luminosity of a few massive spiral galaxies which are measured from $r=(0.05-0.15)r_{\rm 200}$, including firm detection of X-ray emissions from NGC~1961 and NGC~6753. These upper limits on the X-ray luminosities of the extended hot halo will also be compared to the similar X-ray luminosities measured from $r=(0.05-0.15)r_{\rm 200}$ of the CGM-MASS galaxies (Table~\ref{table:LXRadii}). 

In addition to NGC~1961, NGC~6753, and the upper limit of $L_{\rm X, (0.05-0.15)r_{\rm 200}}$ of the galaxies in \citet{Bogdan15}, we also include the Milky Way for comparison. The X-ray luminosity of the Milky Way Galaxy [$(2.0_{-1.2}^{+3.0})\times10^{39}\rm~ergs~s^{-1}$] is obtained from \citet{Snowden97}, while the error range is obtained from \citet{Miller15}. We have assumed most of this X-ray luminosity can be attributed to the hot gas distributed within $0.1r_{\rm 200}$. The temperature ($\approx0.2\rm~keV$) and the $\beta$ index of the radial intensity profile ($0.50\pm0.03$) are also obtained from \citet{Miller15}. The dark matter halo mass [$M_{\rm 200}=(1.79\pm0.16)\times10^{12}\rm~M_\odot$] and virial radius [$r_{\rm 200}=(252.2\pm7.5)\rm~kpc$] are computed from the rotation velocity [$v_{\rm rot}=(218\pm6)\rm~km~s^{-1}$; \citealt{Bovy12}] in the same way as for the CGM-MASS galaxies. The stellar mass [$(6.43\pm0.63)\times10^{10}\rm~M_\odot$] is obtained from \citet{McMillan11}, while the SFR [$(1.065\pm0.385)\rm~M_\odot~yr^{-1}$] is obtained from \citet{Robitaille10}.

\subsection{Massive elliptical galaxies}\label{subsec:EllipticalSample}

Elliptical galaxies have significantly different X-ray scaling relations from disk galaxies over a large mass range (e.g., \citealt{Li13b}). In the present paper, we just \emph{qualitatively} compare the X-ray luminosity and radial distribution (in terms of the $\beta$ index) of the hot gas of the massive spiral galaxies to two samples of massive elliptical galaxies: the MASSIVE sample (\citealt{Ma14,Goulding16}) and \citet{OSullivan03}'s X-ray luminous elliptical galaxy sample. The X-ray luminosity and temperature of hot gas of the MASSIVE sample are directly taken from \citet{Goulding16}, which are extracted within the effective radius of the galaxies (typically $<0.1r_{\rm 200}$ as adopted for the massive spiral galaxies) and are measured in 0.3-5~keV (compared to 0.5-2~keV for spiral galaxies). \citet{OSullivan03}'s sample are based on \emph{ROSAT} observations, so the removal of bright point-like sources may not be as clean as more recent X-ray observations. We just use the $\beta$ index in their sample for a qualitative comparison.


\section{Statistical Analysis}\label{sec:Statistical}

In this section, we statistically compare the hot gas properties to other galaxy properties for the CGM-MASS sample and other galaxy samples as introduced in \S\ref{sec:OtherSample}. As Li14 is the most uniform sample studying the hot gaseous halo of spiral galaxies, many of the comparisons are based on the best-fit relations to a subsample extracted from Li14. We therefore summarize these relations in Table~\ref{table:Statistic}. We caution that although the soft X-ray emission from hot gas around the CGM-MASS galaxies has been detected to $r\sim(30-100)\rm~kpc$ (\S\ref{subsection:SpatialCorona}), the properties of the hot gas at large radii are poorly constrained due to the systematical uncertainties in subtracting the background and the small number of photons. In many of the statistical comparisons presented in this section, we only compare the properties of hot gas measured in the inner halo (except for \S\ref{subsubsec:ScalingRelationOuter} and \ref{subsec:Slope}), which however, are still mainly from the CGM extending out of the galactic disk and bulge (e.g., Fig.~\ref{fig:imagesZoomin}).

\subsection{X-ray scaling relation}\label{subsec:ScalingRelation}

\subsubsection{Scaling relations for hot gas emission from the inner halo}\label{subsubsec:ScalingRelationInner}

\begin{figure*}
\begin{center}
\epsfig{figure=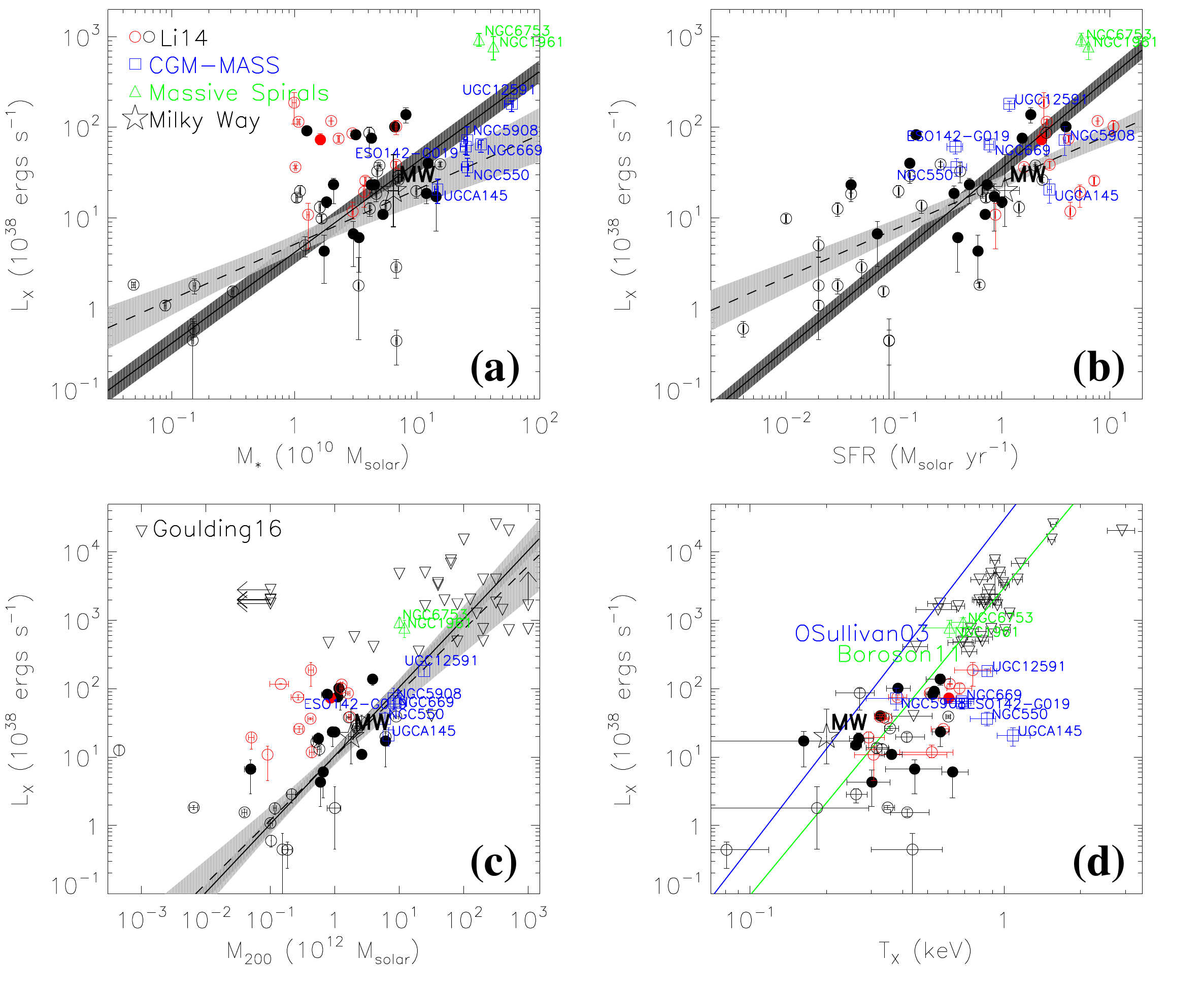,width=1.0\textwidth,angle=0, clip=}
\caption{X-ray scaling relations of the 0.5-2~keV luminosity of the hot gas ($L_{\rm X}$) measured in $r<0.1r_{\rm 200}$ v.s. various galaxy properties: (a) stellar mass ($M_*$); (b) SFR; (c) dark matter halo mass ($M_{\rm 200}$) estimated from the rotation velocity $v_{\rm rot}$; (d) hot gas temperature ($T_{\rm X}$) measured within the spectral analysis region. Symbols representing data from various samples are denoted in panels~(a) and (c). The homogenization of these data are discussed in detail in \S\ref{sec:OtherSample}. Names of the CGM-MASS galaxies, other massive spiral galaxies, and the Milky Way (MW) are denoted beside the data points. The circles (both filled and open) represent the nearby highly-inclined disk galaxies studied in \citet{Li13a,Li13b,Li14}. Red/black circles represent starburst/non-starburst galaxies, while open/filled circles are field/clustered galaxies, respectively. We also include \emph{Chandra} measurements of the MASSIVE early-type galaxy sample \citep{Goulding16} for comparison in panels~(c) and (d) (downward triangles). The blue and green solid lines in panel~(d) are the best-fit $T_{\rm X}-L_{\rm X}$ relations from the high- and low-mass elliptical galaxy samples of \citet{OSullivan03} and \citet{Boroson11}, respectively. The black solid lines in panels~(a-c) are linear fit to different subsamples of \citet{Li13a}'s sample [non-starburst field galaxies in (a); the whole sample in (b); non-starburst field galaxies with $v_{\rm rot}>50\rm~km~s^{-1}$ in (c)] and the dark shaded region represents the 1-$\sigma$ confidence range. The black dashed line and light shaded regions represent non-linear fit (power law with slope set free) and 1-$\sigma$ confidence range of the same subsamples.}\label{fig:scalinginner}
\end{center}
\end{figure*}

In Fig.~\ref{fig:scalinginner}, we present several X-ray scaling relations ($M_*-L_{\rm X}$, ${\rm SFR}-L_{\rm X}$, $M_{\rm 200}-L_{\rm X}$, $T_{\rm X}-L_{\rm X}$) for the hot gas emission from the inner halo (typically $r<0.1r_{\rm 200}$). For X-ray scaling relations between $L_{\rm X}$ and $M_*$, SFR, and $M_{\rm 200}$ (Fig.~\ref{fig:scalinginner}a-c), the CGM-MASS galaxies and the MW are consistent with the non-starburst galaxies in Li14's sample. On the other hand, the two largely face-on and more star formation active massive spiral galaxies NGC~1961 and NGC~6753, appear to be more X-ray luminous on all the scaling relations. We do not rescale the X-ray luminosity of the MASSIVE sample for a uniform comparison, as spatial analysis of the X-ray intensity profile is not presented in \citet{Goulding16}. Since the effective radius of the MASSIVE sample is typically smaller than $0.1r_{\rm 200}$, we expect $L_{\rm X}$ of most of the data points of the MASSIVE galaxies plotted in Fig.~\ref{fig:scalinginner}c are lower limits. Therefore, massive elliptical galaxies are on average more X-ray luminous than spiral galaxies, which is clearer in Fig.~\ref{fig:Xbrightness}c.

\begin{table*}
\vspace{-0.in}
\begin{center}
\caption{Summary of the statistical relations with at least a weak correlation.} 
\footnotesize
\vspace{-0.0in}
\tabcolsep=7.5pt%
\begin{tabular}{lclcccccccccccc}
\hline\hline
Relation        & $r_{\rm s}$ & Fitted relation & Li14 subsample & Figure \\
\hline
$M_*-L_{\rm X}$ & $0.58\pm0.16$ & $L_{\rm X}=(4.15\pm1.18)M_*$ & Non-starburst field galaxy & \ref{fig:scalinginner}a \\
  & - & $L_{\rm X}=(5.11\pm1.21)M_*^{0.61\pm0.14}$ & - & - \\
${\rm SFR}-L_{\rm X}$ & $0.67\pm0.08$ & $L_{\rm X}=(35.9\pm8.4)\rm SFR$ & Removing NGC4342 & \ref{fig:scalinginner}b \\
  & - & $L_{\rm X}=(24.9\pm3.9){\rm SFR}^{0.53\pm0.08}$ & - & - \\
$M_{\rm 200}-L_{\rm X}$ & $0.31\pm0.28$ & $L_{\rm X}=(10.6\pm2.6)M_{\rm 200}$ & Non-starburst field galaxy, $v_{\rm rot}>50\rm~km~s^{-1}$ & \ref{fig:scalinginner}c \\
  & - & $L_{\rm X}=(10.1\pm2.9)M_{\rm 200}^{0.92\pm0.15}$ & - & - \\
$T_{\rm X}-L_{\rm X}$ & $0.43\pm0.15$ & - & All & \ref{fig:scalinginner}d \\
$M_*-L_{\rm X}/M_*$ & $-0.45\pm0.11$ & - & All & \ref{fig:Xbrightness}a \\
${\rm SFR}-L_{\rm X}/M_*$ & $0.36\pm0.11$ & - & All & \ref{fig:Xbrightness}b \\
$M_{\rm 200}-L_{\rm X}/M_{\rm 200}$ & $-0.46\pm0.12$ & - & $v_{\rm rot}>30\rm~km~s^{-1}$ & \ref{fig:Xbrightness}c \\
${\rm SFR}-L_{\rm X}/M_{\rm 200}$ & $0.51\pm0.11$ & - & $v_{\rm rot}>30\rm~km~s^{-1}$ & \ref{fig:Xbrightness}d \\
$\dot{E}_{\rm SN(Ia+CC)}-L_{\rm X}$ & $0.70\pm0.08$ & $L_{\rm X}=(0.81\pm0.12)\dot{E}_{\rm SN(Ia+CC)}$ & Removing NGC4342 & \ref{fig:Ebudget}a \\
  & - & $L_{\rm X}=(1.69\pm0.53)\dot{E}_{\rm SN(Ia+CC)}^{0.76\pm0.08}$ & - & - \\
$M_{\rm TF}/M_*-\eta$ & $0.52\pm0.13$ & $\eta=(0.41\pm0.06)M_{\rm TF}/M_*$ & Removing NGC4438, $v_{\rm rot}>30\rm~km~s^{-1}$ & \ref{fig:Ebudget}b \\
\hline\hline
\end{tabular}\label{table:Statistic}
\end{center}
$r_{\rm s}$ is the Spearman's rank order correlation coefficient. Similar as in \citet{Li13b,Li16a}, we adopt $0.3<|r_{\rm s}|<0.6$ as a weak correlation, and $|r_{\rm s}|>0.6$ as a tight correlation, with negative $r_{\rm s}$ representing anti-correlation. Unit of the parameters in the ``Fitted relation'' column are presented on the related figures. For some relations, we have presented fitting with both linear and non-linear models. $r_{\rm s}$ and the fitted relation are obtained basically based on \citet{Li14}'s sample, but the real adopted subsample has been slightly modified as indicated in the ``Li14 subsample'' column. In this column, ``All'' means all the sample galaxies with a well estimate of the related parameters, while in some cases, a lot of galaxies have been removed from the calculation because the parameters are not well constrained (e.g., $T_{\rm X}$).
\\
\end{table*}

In order to create fiducial relations for comparison, we also fit the $M_*-L_{\rm X}$, ${\rm SFR}-L_{\rm X}$, and $M_{\rm 200}-L_{\rm X}$ relations for some subsamples of Li14's sample. In particular, for the $M_*-L_{\rm X}$ relation, only non-starburst field galaxies (open black circles) are included in the fitting, as starburst or clustered galaxies appear to be systematically more X-ray luminous. For the ${\rm SFR}-L_{\rm X}$ relation, we include all the galaxies in Li14's sample to expand the range of SFR, although starburst galaxies may be slightly less luminous in X-ray at a given SFR. Similar as for the $M_*-L_{\rm X}$ relation, we only include non-starburst field galaxies in the fitting of the $M_{\rm 200}-L_{\rm X}$ relation. We further remove galaxies with $v_{\rm rot}\leq50\rm~km~s^{-1}$. These galaxies are often interacting systems, so the estimate of $M_{\rm 200}$ based on $v_{\rm rot}$ may not be correct. 

We fit the selected galaxies with both a linear model (solid line) and a non-linear model (dashed line), following the method described in \citet{Li13b}. In order to estimate the errors of the fitted parameters, we first generate 1000 bootstrap-with-replacement samples of the data points from the selected subsamples and then resample each data point, assuming a normal distribution with the expected value and errors. For each re-generated sample, we fit the data with the same expression to obtain its parameters. These measurements are then rank ordered; their 68 per cent percentiles around the median fitting parameters (taken as the best-fit parameters) are taken as their 1~$\sigma$ uncertainties, which account for the systematic dispersion among the original data points as well as the uncertainties in their individual measurements. The 1~$\sigma$ uncertainties of the fitted relations are shown as shaded areas (dark for linear model, light for non-linear model). We caution that since the linear model does not account for the variation of the slope, the 1~$\sigma$ uncertainty just includes the variation of the ``best-fit'' normalization in the fitting of each resampled dataset. Therefore, the error does not reflect the real scatter of the data.

Comparison with the fiducial best-fit relation confirms our previous argument that the CGM-MASS galaxies and the MW are consistent with lower mass galaxies on the X-ray scaling relations, but NGC~1961 and NGC~6753 are more X-ray luminous. Specifically, the measured or average luminosity of the CGM-MASS galaxies/MW/NGC1961/NGC6753 are -0.30/-0.12/+0.65/+0.85~dex from the best-fit linear $M_*-L_{\rm X}$ relation, and +0.18/+0.10/+1.20/+1.35~dex from the best-fit non-linear $M_*-L_{\rm X}$ relation. For comparison, the 1~$\sigma$ scatter of the data points included in the fitting around the best-fit linear relation is 0.85~dex. The other relations have similar behaviour: for the ${\rm SFR}-L_{\rm X}$ relation, the CGM-MASS/MW/NGC1961/NGC6753 are +0.19/-0.28/+0.54/+0.68~dex from the best-fit linear relation, and +0.36/-0.11/+1.07/+1.19~dex from the best-fit non-linear relation, with a 1~$\sigma$ scatter of the data points of 1.69~dex; for the $M_{\rm 200}-L_{\rm X}$ relation, the CGM-MASS/MW/NGC1961/NGC6753 are -0.22/+0.02/+0.79/+0.95~dex from the best-fit linear relation, and -0.12/+0.06/+0.90/+1.05~dex from the best-fit non-linear relation, with a 1~$\sigma$ scatter of the data points of 1.14~dex.

Early-type galaxies, especially massive ones, often have well defined scaling relations between the galaxy mass, hot gas luminosity and temperature. In Fig.~\ref{fig:scalinginner}d, we compare spiral galaxies from the CGM-MASS and Li14's samples to the MASSIVE sample \citep{Goulding16} and the best-fit $T_{\rm X}-L_{\rm X}$ relations from \citet{OSullivan03} and \citet{Boroson11} for massive and dwarf elliptical galaxies, respectively. Although the MW, NGC~1961, NGC~6753, and some of Li14's sample fall on the relationships defined by elliptical galaxies, most of the CGM-MASS galaxies (except for NGC~5908) and some galaxies in Li14's sample have higher hot gas temperature at a given X-ray luminosity, probably indicating a systematical bias from the well defined scaling relations. 

\subsubsection{Scaling relations for the specific properties of the hot halo}\label{subsubsec:ScalingSpecific}

\begin{figure*}
\begin{center}
\epsfig{figure=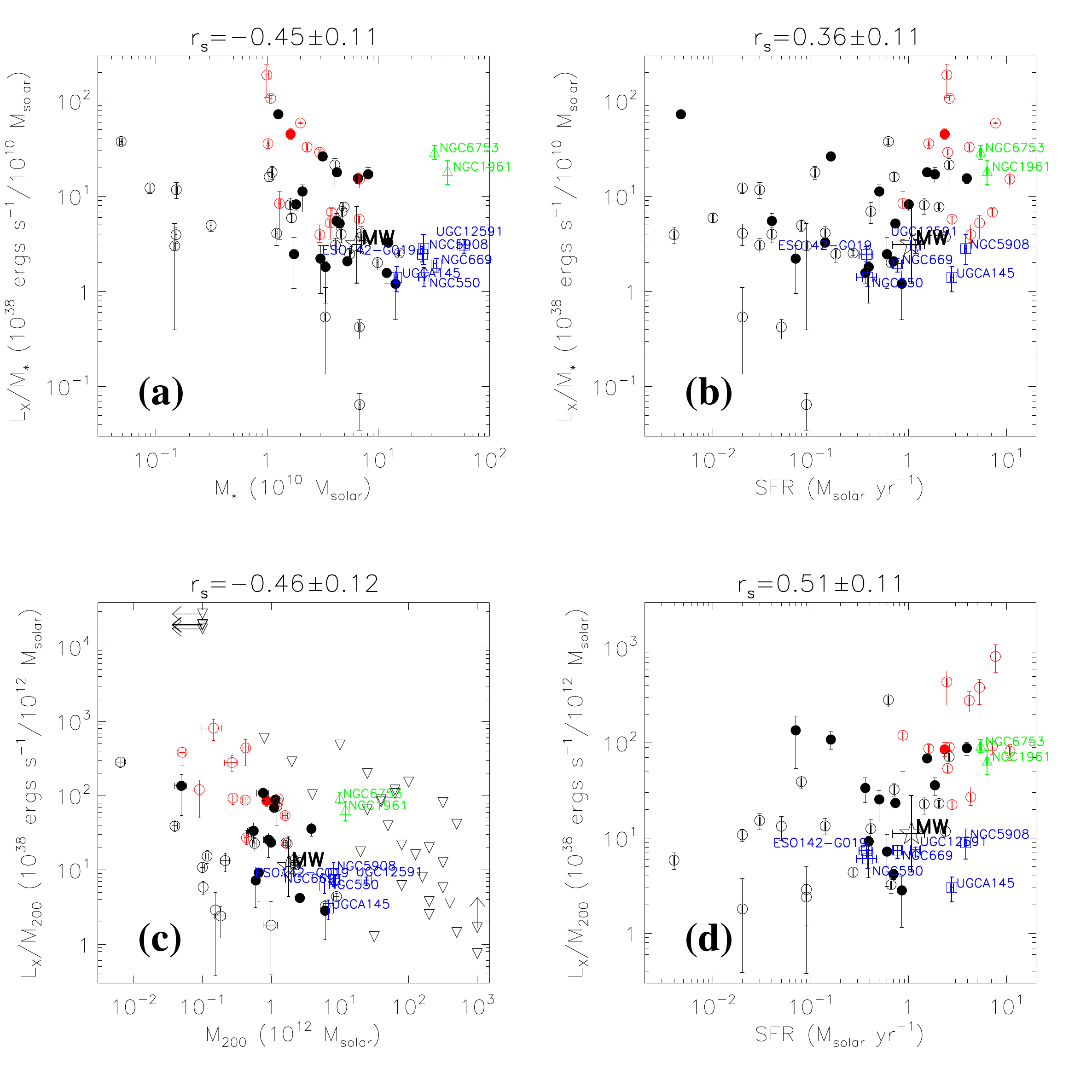,width=1.0\textwidth,angle=0, clip=}
\caption{Specific X-ray luminosity ($L_{\rm X}/M_*$ or $L_{\rm X}/M_{\rm 200}$) plotted against other galaxy properties. The Spearman's rank order correlation coefficient $r_{\rm s}$ are calculated for all the galaxies plotted in each panel, except for the MASSIVE sample in panel~(c) (open downward triangles).}\label{fig:Xbrightness}
\end{center}
\end{figure*}

Scaling relations in Fig.~\ref{fig:scalinginner} are for the integrated properties of galaxies, which could be affected by a general scaling of galaxies, i.e., bigger galaxies tend to have higher stellar mass, SFR, X-ray luminosity, and hot gas temperature (e.g., \citealt{Wang16}). We therefore present scaling relations for some specific properties of galaxies in Fig.~\ref{fig:Xbrightness} ($M_*-L_{\rm X}/M_*$, ${\rm SFR}-L_{\rm X}/M_*$, $M_{\rm 200}-L_{\rm X}/M_{\rm 200}$, ${\rm SFR}-L_{\rm X}/M_{\rm 200}$). 

We herein quantify the goodness of correlation of some subsamples from Li14 with the Spearman's rank order coefficient ($r_{\rm s}$, which is shown on top of each panel in Fig.~\ref{fig:Xbrightness} and summarized in Table~\ref{table:Statistic}). Following \citet{Li13b}, we consider $|r_{\rm s}|\gtrsim0.6$, $0.6\gtrsim|r_{\rm s}|\gtrsim0.3$, and $|r_{\rm s}|\lesssim0.3$ as tight, weak, and no correlations. Positive or negative values of $r_{\rm s}$ indicate positive or anti-correlations.

As shown in Fig.~\ref{fig:Xbrightness}a,c, there are weak negative correlations between the specific X-ray luminosity ($L_{\rm X}/M_*$ or $L_{\rm X}/M_{\rm 200}$) and galaxy or halo mass (the MASSIVE galaxies are just plotted for a qualitative comparison so are not included in calculating $r_{\rm s}$). These negative correlations indicate the $M_*-L_{\rm X}$ or $M_{\rm 200}-L_{\rm X}$ relations must be sublinear, which is not clearly indicated in Fig.~\ref{fig:scalinginner}a,c due to the large scatter of the data points. In comparison, similar mass-$L_{\rm X}$ relationships (mass often expressed in optical or near-IR luminosity) of early-type galaxies are often superlinear (the logarithm slope is typically $\gtrsim2$), and the relations for galaxy groups/clusters are even steeper (e.g., \citealt{Ponman99,OSullivan03,Boroson11,Li11}). This trend is also indicated by the higher $L_{\rm X}/M_{\rm 200}$ of MASSIVE galaxies than those of spiral galaxies, especially when $L_{\rm X}$ of the MASSIVE galaxies are measured at smaller radii (\S\ref{subsec:EllipticalSample}). 

The steeper mass-$L_{\rm X}$ relations of more massive elliptical galaxies and groups/clusters of galaxies are a result of strong gravitational heating and confinement, which do not seem to be quite important for most of the low mass spiral galaxies. There are not enough massive spiral galaxies (e.g., with $\log M_*\gtrsim11$ or  $\log M_{\rm 200}\gtrsim12.5$) for us to claim for any possible variations of the slope of the ${\rm mass}-L_{\rm X}$ relation with the mass of the galaxies for spiral galaxies only. More X-ray observations of massive spiral galaxies are needed to further examine such a trend.

On the other hand, the specific X-ray luminosity has a weak positive correlation with the SFR for most of the spiral galaxies (Fig.~\ref{fig:Xbrightness}b,d). All the massive spiral galaxies, including the CGM-MASS sample, MW, NGC~1961, and NGC~6753, are just marginally consistent with these relations, and appear to be the least X-ray luminous at a given SFR.

\subsubsection{Scaling relations for hot gas emission from the outer halo}\label{subsubsec:ScalingRelationOuter}

\begin{figure}
\begin{center}
\epsfig{figure=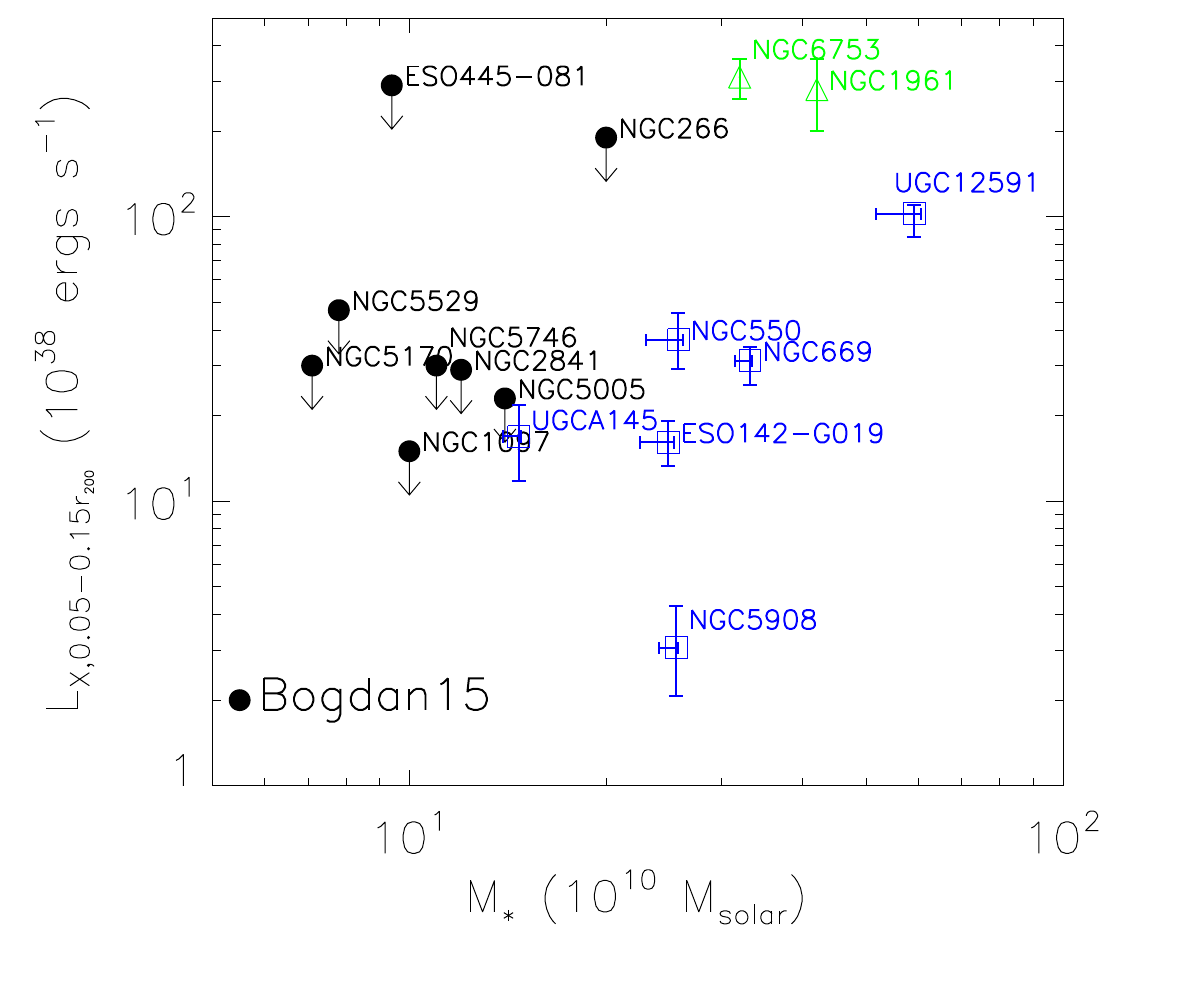,width=0.5\textwidth,angle=0, clip=}
\caption{0.5-2~keV luminosity measured in $r=(0.05-0.15)r_{\rm 200}$ ($L_{\rm X,0.05-0.15r_{200}}$) v.s. stellar mass ($M_*$) of the CGM-MASS galaxies (blue boxes) and the massive spiral galaxies in \citet{Bogdan15}'s sample (green triangles and black circles). All the galaxies in \citet{Bogdan15}, except for NGC~1961 and NGC~6753, have just upper limit constraint on the X-ray luminosity measured in this radial range. $L_{\rm X,0.05-0.15r_{200}}$ of the CGM-MASS galaxies are estimated based on the luminosity measured in the spectral analysis region and the best-fit radial intensity profile (Table~\ref{table:LXRadii}).}\label{fig:scalingouter}
\end{center}
\end{figure}

We compare the CGM-MASS galaxies to the massive spiral galaxies of \citet{Bogdan15} (including NGC~1961 and NGC~6753) on the $M_*-L_{\rm X}$ relation for the soft X-ray luminosity measured in $r=(0.05-0.15)r_{\rm 200}$ (Fig.~\ref{fig:scalingouter}). Most of \citet{Bogdan15}'s sample galaxies do not have extended X-ray emission detected at such large radii, so most of the data points in Fig.~\ref{fig:scalingouter} are upper limits on $L_{\rm X}$. The stellar mass of \citet{Bogdan15}'s sample is $M_*\approx(0.7-2.0)\times10^{11}\rm~M_\odot$, while the SFR is in the range of $(0.4-5.8)\rm~M_\odot~yr^{-1}$.  Therefore, most of \citet{Bogdan15}'s sample galaxies are extremely quiescent in star formation, and none of them can be regarded as starburst galaxies according to \citet{Li13a}'s criteria (typically equals to ${\rm SFR}/M_*\gtrsim1\rm~M_\odot~yr^{-1}/10^{10}\rm~M_\odot$; see Fig.~1d of \citealt{Li13b}).

For these massive quiescent spiral galaxies, X-ray emission at such large galactocentric radii is expected to be produced by an extended corona not directly related to current star formation feedback. If the extended X-ray emission is mainly produced by the gravitationally heated virialized gas, similar as in more massive systems such as galaxy clusters, we expect there is a correlation between $L_{\rm X}$ and $M_*$ or $M_{\rm 200}$. However, the mass range of the galaxies with a clear detection of the extended X-ray emission is too narrow, and many of the X-ray measurements are just upper limit constraint on $L_{\rm X}$. We examined similar scaling relations as for the inner halo (Figs.~\ref{fig:scalinginner}, \ref{fig:Xbrightness}), but do not find any significant correlations. From Fig.~\ref{fig:scalingouter}, we can only conclude that the current data do not conflict with the hypothesis that the gas is gravitationally heated and there is a positive correlation between $L_{\rm X}$ and $M_*$.

\subsection{Temperature of the hot halo gas}\label{subsec:temperature}

We compare the measured hot gas temperature in the inner halo to the virial temperature of the galaxies. We estimate the virial temperature $T_{\rm virial}$ of the galaxies within $r_{\rm 200}$:
\begin{equation}\label{equi:Tvirial}
T_{\rm virial}=\frac{2}{3}\frac{GM_{\rm 200}}{r_{\rm 200}}\frac{\mu m_{\rm H}}{k_{\rm B}},
\end{equation}
where $\mu$ is the mean atomic weight of the gas, $m_{\rm H}$ is the mass of the hydrogen atom, $G$ is the gravitational constant, and $k_{\rm B}$ is the Boltzmann constant. We assume the hot gas metallicity to be $0.2\rm~Z_\odot$ for all the galaxies to estimate the mean atomic weight $\mu$. The assumption on gas metallicity does not affect the result significantly. The current data do not provide any strict constraint on the radial variation of the hot gas temperature, so we simply assume there is no radial variation.

\begin{figure}
\begin{center}
\epsfig{figure=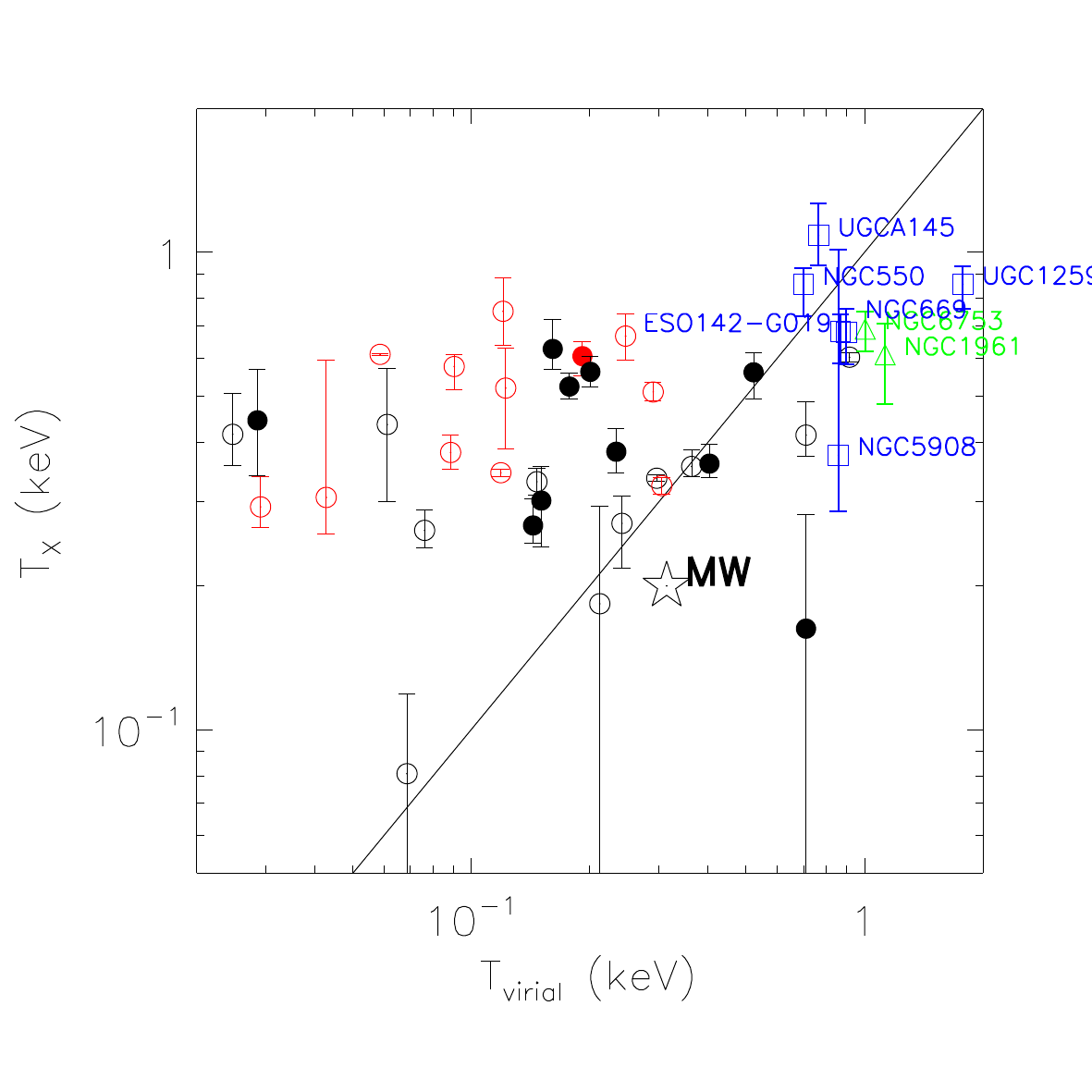,width=0.5\textwidth,angle=0, clip=}
\caption{Hot gas temperature ($T_{\rm X}$) measured within the spectral analysis region v.s. the virial temperature of the galaxies ($T_{\rm virial}$) estimated from their rotation velocity ($v_{\rm rot}$). Symbols are the same as in Fig.~\ref{fig:scalinginner}. The solid line indicates where $T_{\rm X}=T_{\rm virial}$.}\label{fig:VirialTemperature}
\end{center}
\end{figure}

$T_{\rm virial}$ is plotted against the measured hot gas temperature in the inner halo in Fig.~\ref{fig:VirialTemperature}. Most of Li14's sample galaxies, especially starburst ones, have hot gas temperature significantly higher than the virial temperature, so the gas must be heated by non-gravitational processes. However, some of the non-starburst galaxies in Li14's sample and all of the massive spiral galaxies (CGM-MASS, MW, NGC~1961, NGC~6753) have hot gas temperature comparable to the virial temperature. The uncertainty of hot gas temperature measurement for X-ray faint galaxies is very large and the temperature is only estimated at much smaller radii than $r_{\rm 200}$. Furthermore, temperature of the virialized gas in low mass galaxies may fall below that of the X-ray emitting range. All these uncertainties may bias the comparison in Fig.~\ref{fig:VirialTemperature}, especially at the low mass end. However, the consistency between the measured hot gas temperature and the virial temperature in massive galaxies, as well as the significant difference between massive and lower mass (especially starburst ones) spiral galaxies on the $T_{\rm virial}-T_{\rm X}$ plot, strongly indicate that gravitational processes can be important in the heating and dynamics of the gas in these extremely massive isolated spiral galaxies.

\subsection{Slope of the radial X-ray intensity profile}\label{subsec:Slope}

\begin{figure*}
\begin{center}
\epsfig{figure=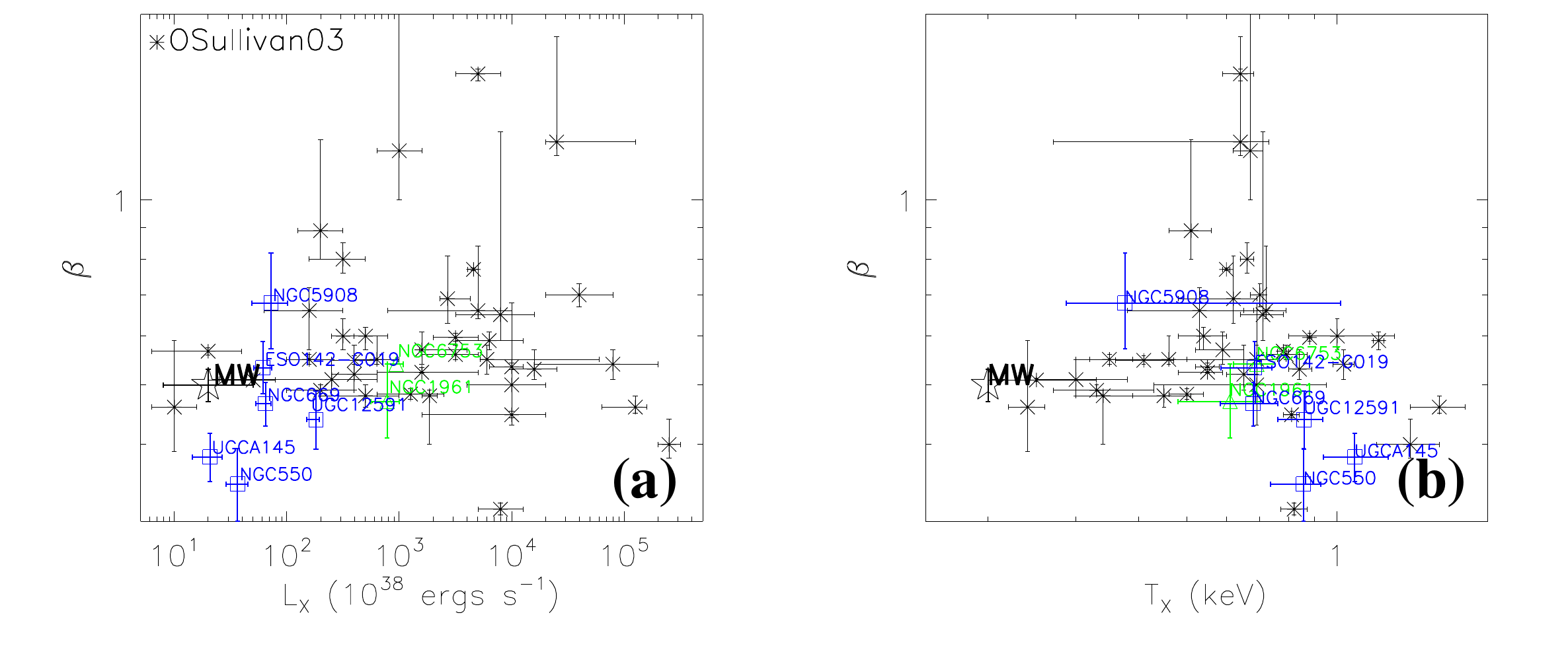,width=1.0\textwidth,angle=0, clip=}
\caption{$\beta$-index of the radial intensity distribution of the hot gas component (described with a $\beta$ function) v.s. various galaxy properties. $L_{\rm X}$ of the CGM-MASS galaxies, NGC~1961, NGC~6753, and the MW in panel~(c) are measured in $r<0.1r_{\rm 200}$, while $T_{\rm X}$ in panel~(d) are measured in the spectral analysis regions presented in Fig.~\ref{fig:imagesZoomin}. We also include \citet{OSullivan03}'s sample of elliptical galaxies in panels~(c) and (d) for comparison. Parameters of this sample, however, are not obtained in a uniform way as other galaxy samples, so are just plotted here for a qualitative comparison.}\label{fig:slope}
\end{center}
\end{figure*}

We examine the slope ($\beta$ index) of the radial soft X-ray intensity profile of different galaxy samples in Fig.~\ref{fig:slope}. In order to enlarge the range of galaxy parameters, we include \citet{OSullivan03}'s sample of X-ray luminous elliptical galaxies in Fig.~\ref{fig:slope}c,d for comparison. These elliptical galaxies may have quite different formation histories and hot halo properties from the spiral galaxies studied in this paper, and the X-ray properties are measured in different ways (from \emph{ROSAT} observations, at different radii, and stellar contributions are not well accounted for). Therefore, the comparison is just qualitative.

In general, we do not find any significant correlations between $\beta$ and other galaxy parameters for the massive spiral galaxies (the CGM-MASS galaxies, NGC~1961, NGC~6753, the MW; Fig.~\ref{fig:slope}). Most of these massive spiral galaxies have $\beta$ in a narrow range of $\approx0.35-0.55$, except for the relatively large value of NGC~5908, which is largely uncertain due to the removal of the X-ray bright AGN (Figs.~\ref{fig:imagesZoomin}, \ref{fig:RadialProfile}; Paper~I). In addition, the systematical uncertainty in subtracting the stellar and background components may also affect the fitted value of $\beta$. Within the large statistical error shown in Fig.~\ref{fig:slope} and these systematical errors, the $\beta$ indexes of massive spiral galaxies are consistent with those of X-ray luminous elliptical galaxies at a given hot gas X-ray luminosity or temperature (Fig.~\ref{fig:slope}).

\subsection{Energy budget of galactic corona}\label{subsec:EnergyBudget}

\begin{figure*}
\begin{center}
\epsfig{figure=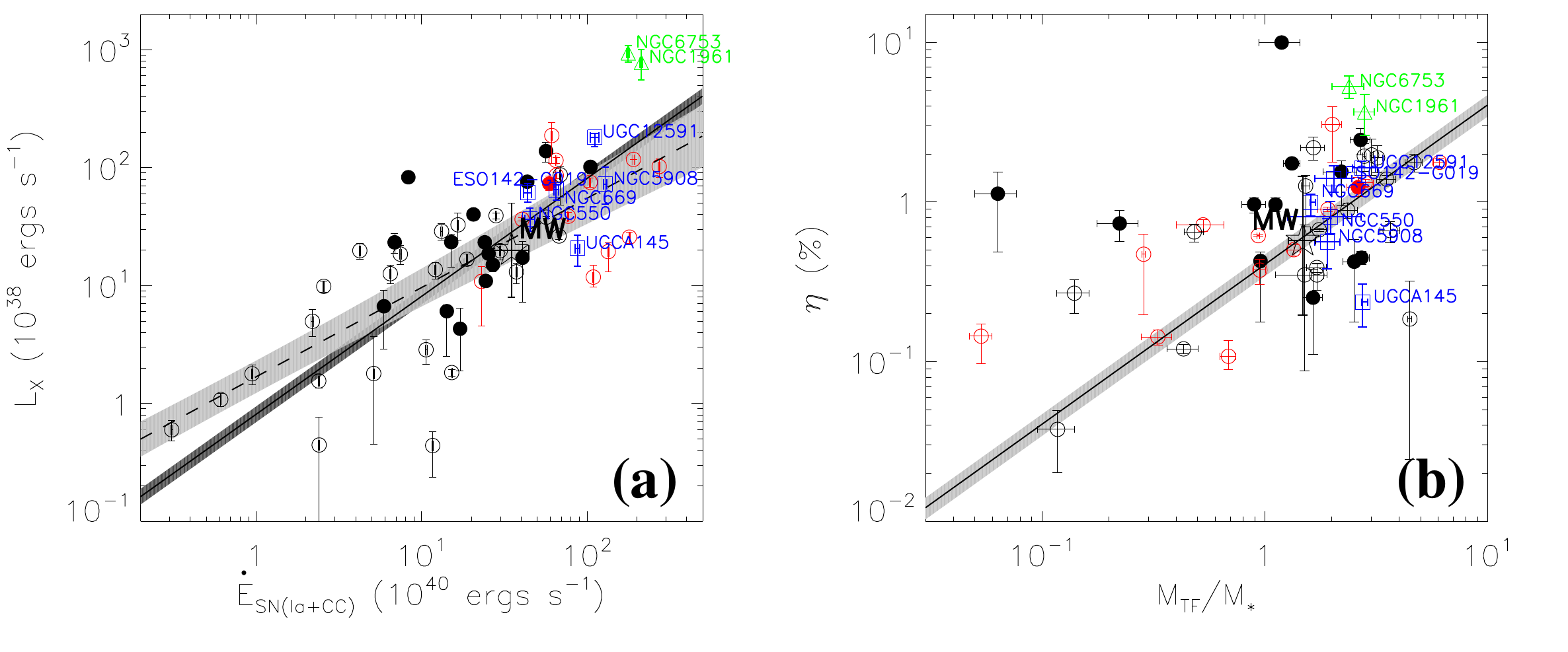,width=1.0\textwidth,angle=0, clip=}
\caption{(a) $L_{\rm X}$ measured in $r<0.1r_{\rm 200}$ v.s. the total (Type~Ia + core collapsed) SN energy injection rate [$\dot{E}_{\rm SN(Ia+CC)}$]. (b) The X-ray radiation efficiency [$\eta\equiv L_{\rm X}/\dot{E}_{\rm SN(Ia+CC)}$] v.s. the baryonic to stellar mass ratio ($M_{\rm TF}/M_*$). Symbols are the same as in Fig.~\ref{fig:scalinginner}.}\label{fig:Ebudget}
\end{center}
\end{figure*}

Following the same method as adopted in \citet{Li13b,Li16a}, we convert the stellar mass of the galaxies to the Type~Ia SNe energy injection rate and the SFR to the core collapsed (CC) SNe energy injection rate, in order to examine the energy budget of the galactic corona. The $\dot{E}_{\rm SN}-L_{\rm X}$ relation is presented in Fig.~\ref{fig:Ebudget} ($\dot{E}_{\rm SN}$ is the total energy injection rate by Ia+CC SNe). Similar as the other scaling relations, the CGM-MASS galaxies and the MW are consistent with Li14's lower mass galaxies on the $\dot{E}_{\rm SN}-L_{\rm X}$ relation (CGM-MASS galaxies and the MW are -0.01dex and -0.15dex from the best-fit linear relation; the scatter around the best-fit relation is 0.49~dex), indicating a small fraction (typically $\lesssim1\%$; \citealt{Li13b}) of SNe energy has been converted to soft X-ray emission. In comparison, NGC~1961/NGC~6753 are 0.66/0.82~dex above the best-fit linear $\dot{E}_{\rm SN}-L_{\rm X}$ relation, which is significantly larger than most other galaxies including the starburst ones, except for a few clustered galaxies whose X-ray emission may be contaminated by the ICM.

The difference of the coronal gas energy budget between the CGM-MASS/MW galaxies and NGC~1961/6753 are more clearly shown through the $M_{\rm TF}/M_*-\eta$ relation (Fig.~\ref{fig:Ebudget}b). $M_{\rm TF}/M_*$ is the dynamical to photometry mass ratio of the galaxy ($M_*$ is obtained from K-band luminosity, while $M_{\rm TF}$ is obtained from the rotational velocity; \S\ref{subsubsec:OtherProperty}), and $\eta$ is the X-ray radiation efficiency defined as $\eta\equiv L_{\rm X}/\dot{E}_{\rm SN(Ia+CC)}$. A tight correlation between $M_{\rm TF}/M_*$ and $\eta$ is found in \citet{Li13b}. The correlation has been explained as a combination effect of gravitational confinement (proportional to $M_{\rm TF}$) and the heating of the gas via galactic feedback (related to $M_*$), which have opposite effects on the X-ray emissivity in the inner halo of the galaxies. It is clear that all the massive spiral galaxies (CGM-MASS, MW, NGC~1961, and NGC~6753) have similar $M_{\rm TF}/M_*$, but $\eta$ differs by a factor of $\sim30$, with the CGM-MASS and MW galaxies having $\eta\approx(0.2-2)\%$ and consistent with lower mass field galaxies from Li14, while NGC~1961/6753 have $\eta\sim5\%$.

\section{Discussion}\label{section:discussion}

\subsection{Evidence for the presence of gravitational heating of the hot halo gas}\label{subsection:GravitationalHeatingGas}

There are in general two major heating sources of the hot halo gas, either from gravitational processes (shock or compression) or from various forms of galactic feedback. The tight correlation between $L_{\rm X}$ and SFR or $\dot{E}_{\rm SN}$, as well as some specific connections of extraplanar hot gas features with disk star formation regions revealed in previous works (e.g., \citealt{Li08,Li13a}), strongly indicate the halo X-ray emission is at least partly related to the feedback from stellar sources, if not all produced by them. We then first investigate if gravitational heating could possibly contribute in producing the hot gas.

The escaping velocity of a galaxy determines whether the galactic outflow could escape into the intergalactic space or be thermalized locally within the gravitational potential of the dark matter halo. The escaping velocity at the edge of the galactic disk can be estimated from the circular velocity of the galaxy ($V_{\rm c}$) in the form of \citep{Benson00}: 
\begin{equation}\label{equi:vesc}
v_{\rm esc}=V_{\rm c}[2\ln(r_{\rm vir}/r_{\rm disk})+2]^{1/2}. 
\end{equation}
Assuming $V_{\rm c}=v_{\rm rot}$ and $r_{\rm vir}=r_{\rm 200}$, we can adopt typical parameters of the CGM-MASS galaxies to estimate their escaping velocity.
Adopting a rotation velocity of $v_{\rm rot}=350\rm~km~s^{-1}$ (the corresponding $r_{\rm 200}\approx420\rm~kpc$) and a galactic disk radius of $r_{\rm disk}=20\rm~kpc$, we obtain $v_{\rm esc}\approx10^3\rm~km~s^{-1}$. In comparison, a MW sized galaxy ($v_{\rm rot}=218\rm~km~s^{-1}$, $r_{\rm disk}=15\rm~kpc$) has $v_{\rm esc}\approx6\times10^2\rm~km~s^{-1}$ and most of Li14's sample galaxies should have $v_{\rm esc}\lesssim5\times10^2\rm~km~s^{-1}$. In a typical galactic superwind, most of the X-ray emitting gas has a velocity of $>5\times10^2\rm~km~s^{-1}$ (e.g., \citealt{Strickland00}), so they can escape out of most of the galaxies in Li14's sample. However, it is very likely that the hot gas outflow could not escape out of a galaxy as massive as the CGM-MASS galaxies, especially when the SFR is too low to drive a galactic superwind (the velocity is typically $<5\times10^2\rm~km~s^{-1}$ for a subsonic outflow at low SFR; e.g., \citealt{Tang09}). If the bulk of the hot gas outflow cannot escape, we would expect some of the gas is thermalized and confined within the galactic halo. In this case, gravitational processes could contribute to the heating of the halo gas.

We further search for signatures of gravitational heating from the radial distribution of hot gas emission, which could be affected by many factors, such as the density profile of the dark matter halo, the galactic feedback, and the metal enrichment, etc. If the hot gas around galaxies is in a hydrostatic state and is isothermal, the gas density distribution can be described by a King profile \citep{Cavaliere76}, which naturally produces a $\beta$-function distribution of the radial X-ray intensity profile (Equ.~\ref{equi:Ibeta}; see discussions in \citealt{Jones84}). The $\beta$ index is linked to the energy density ratio of the gravitational energy and hot gas thermal energy in the form of \citep{Jones84}:
\begin{equation}\label{equi:betapredict}
\beta_{\rm predict}=\mu m_{\rm H}\sigma_{\rm v}^2/3k_{\rm B}T_{\rm X},
\end{equation}
where $\sigma_{\rm v}$ is the velocity dispersion of the galaxy and $T_{\rm X}$ is the temperature of the hot gas. If there are additional heating sources such as shock heating from galactic feedback, the radial X-ray intensity distribution is expected to be shallower (smaller $\beta$ index). This is supported by the shallower X-ray intensity profile of galaxy groups and clusters with decreasing hot gas temperature (e.g., \citealt{Ponman99}).

\begin{figure}
\begin{center}
\epsfig{figure=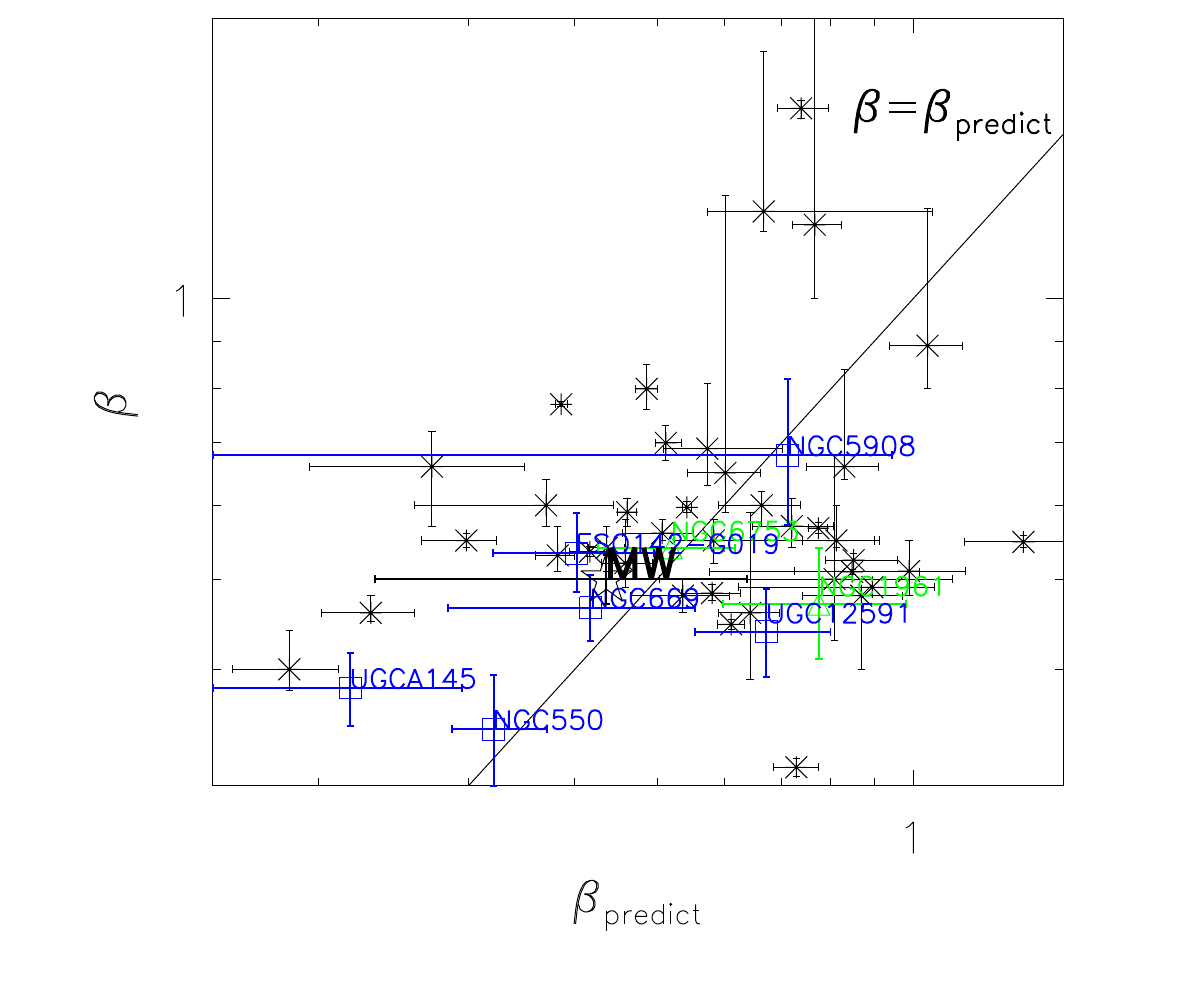,width=0.5\textwidth,angle=0, clip=}
\caption{The measured and predicted $\beta$ index of the $\beta$ function describing the radial intensity profiles of massive spiral galaxies and \citet{OSullivan03}'s X-ray luminous elliptical galaxies. $\beta_{\rm predict}$ is estimated by assuming a hydrostatic-isothermal model of the hot halo gas \citep{Jones84}. See the text for details. Symbols are the same as in Fig.~\ref{fig:slope}. The solid line indicates where $\beta=\beta_{\rm predict}$.}\label{fig:betabetapredict}
\end{center}
\end{figure}

In order to investigate if the gravitation of the galaxy plays a key role in shaping the radial distribution of hot gas, we estimate the predicted $\beta$ index ($\beta_{\rm predict}$) from the measured hot gas temperature and the rotation velocity of the galaxies using Eq.~\ref{equi:betapredict}. In order to convert the measured $v_{\rm rot}$ of massive spiral galaxies to the velocity dispersion $\sigma_{\rm v}$ (\citealt{OSullivan03}'s sample have $\sigma_{\rm v}$ listed in the paper), we adopt the observed linear relation between $\sigma_{\rm v}$ and the circular velocity ($V_{\rm c}$) of a sample of disk and elliptical galaxies from \citet{Corsini05}. Similar as above, we neglect the difference between $v_{\rm rot}$ and $V_{\rm c}$ in this conversion. We also assume a hot gas metallicity of $0.2\rm~Z_\odot$ for both the massive spiral galaxies and \citet{OSullivan03}'s sample, in order to estimate the mean atomic weight $\mu$. 

The estimated $\beta_{\rm predict}$ is compared to the measured $\beta$ in Fig.~\ref{fig:betabetapredict}. The correlation is not significant partially due to the large error of both $\beta$ and $\beta_{\rm predict}$ (only statistical error is included in the plot), but all the galaxies have the measured $\beta$ at least not inconsistent with the predicted $\beta_{\rm predict}$, indicating that massive spiral galaxies and X-ray bright elliptical galaxies apparently have radial distribution of hot gas shaped by similar processes. $\beta\approx\beta_{\rm predict}$ also suggests a hydrostatic isothermal hot gas halo. Although such a dynamical state is not well constrained with the current measurement of $\beta$, it is consistent with the deep gravitational potential of the CGM-MASS galaxies and the discussions in the next two sections (\S\ref{subsection:Thermodynamics},\ref{subsection:MissingFeedback}).

\subsection{Thermodynamics of the CGM-MASS galaxies}\label{subsection:Thermodynamics}

In this section, we investigate the thermodynamics of the hot halo gas of the CGM-MASS galaxies by comparing the radiative cooling ($t_{\rm cool}$) and free fall timescales ($t_{ff}$) of the halo gas.
It has been suggested that the thermodynamics of the hot atmosphere and the presence of multi-phase gas (both hot gas and cool gas) around both massive elliptical galaxies and galaxy clusters are strongly dependent on the ratio between the cooling time and the free fall time ($t_{\rm cool}/t_{ff}$; \citealt{Voit15a,Voit15b}). 

We first estimate the free fall timescale of a cold gas cloud at a distance of $r_{\rm cloud}$ from the galactic center:
\begin{equation}\label{equi:tff}
t_{ff}=(2r_{\rm cloud}/g)^{1/2}=(\frac{2r_{\rm cloud}^3}{GM_{\rm tot}})^{1/2},
\end{equation}
where $g$ is the local gravitational acceleration at $r_{\rm cloud}$ and $M_{\rm tot}$ is the total gravitational mass enclosed by $r_{\rm cloud}$. For simplicity, we have assumed only the mass enclosed by $r_{\rm cloud}$ could affect the dynamics of the gas. Assuming the dark matter halo has a NFW density profile \citep{Navarro97} in the form of:
\begin{equation}\label{equi:NFW}
\rho(r)=\frac{4\rho_{\rm s}}{(r/r_{\rm s})(1+r/r_{\rm s})^2},
\end{equation}
where $r_{\rm s}$ is a characteristic scale radius defined by the virial radius $r_{\rm 200}$ and the concentration factor $c$ as: $r_{\rm s}=r_{\rm 200}/c$. Integrating Equ.~\ref{equi:NFW}, we can derive $\rho_{\rm s}$ with the halo mass $M_{\rm 200}$:
\begin{equation}\label{equi:rhos}
\rho_{\rm s}=\frac{M_{\rm 200}}{16\pi(r_{\rm 200}/c)^3}[\ln(c+1)+\frac{1}{c+1}-1]^{-1}.
\end{equation}

\begin{figure}
\begin{center}
\epsfig{figure=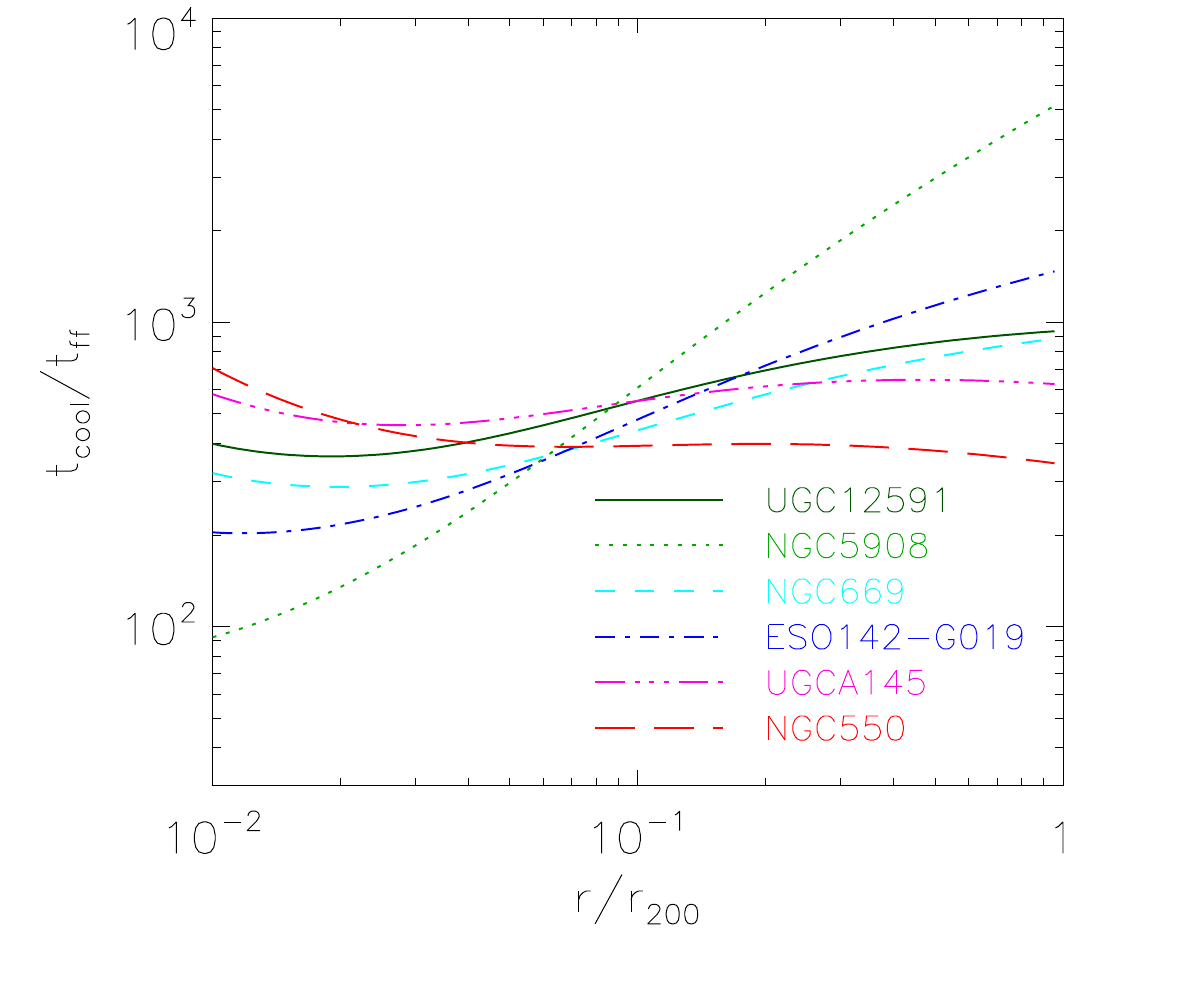,width=0.5\textwidth,angle=0, clip=}
\caption{Radial profile of $t_{\rm cool}/t_{ff}$ of different CGM-MASS galaxies. Note that we have assumed the stellar mass of the galaxy is a mass point, which is only valid at large radii. Therefore, at small radii (typically $<0.1r_{\rm 200}$), the real $t_{\rm cool}/t_{ff}$ should be smaller than plotted on the figure.}\label{fig:tcooltffprofile}
\end{center}
\end{figure}

Substituting Equ.~\ref{equi:NFW} and \ref{equi:rhos} into Equ.~\ref{equi:tff}, we obtain $t_{ff}$ at a given radius $r$ as:
\begin{equation}\label{equi:tffcalc}
t_{ff}=(\frac{2r^3}{GM_{\rm 200}})^{1/2}[\frac{\ln(\frac{cr}{r_{\rm 200}}+1)+\frac{1}{\frac{cr}{r_{\rm 200}}+1}-1}{\ln(c+1)+\frac{1}{c+1}-1}+\frac{M_*}{M_{\rm 200}}]^{-
\frac{1}{2}}.
\end{equation}
We have assumed the stellar content of the galaxy is a point source in deriving the above equation, so it is only valid at large enough radii enclosing most of the stellar mass of the galaxy. Assuming a typical concentration factor of $c=10$ and adopting $M_*$ (Table~\ref{table:GalaxyPara}), $M_{\rm 200}$ and $r_{\rm 200}$ of the CGM-MASS galaxies (Paper~I), we calculate $t_{\rm cool}/t_{ff}$ using Equ.~\ref{equi:tcoolbeta} and \ref{equi:tffcalc} and plot it against $r/r_{\rm 200}$ in Fig.~\ref{fig:tcooltffprofile}.

Only at small enough $t_{\rm cool}/t_{ff}$, radiative cooling is efficient for some of the hot CGM to condense into cold clouds and precipitate onto the galactic disk.
A commonly adopted criterion is $t_{\rm cool}/t_{ff}\lesssim10$ (e.g., \citealt{Voit15a,Voit15b}). As shown in Fig.~\ref{fig:tcooltffprofile}, $t_{\rm cool}/t_{ff}>>10$ throughout the halo for all the CGM-MASS galaxies. A similar result is also found in NGC~1961 \citep{Anderson16}. Therefore, the accretion of cold clouds condensed from the hot halo is not an important source of the star formation fuel for the CGM-MASS galaxies. Most of the hot CGM cannot fall back to the galactic disk in such massive spiral galaxies. This is consistent with their extremely low cold gas content and SFR. 

The above criterion of $t_{\rm cool}/t_{ff}$ is based on the precipitation scenario developed in \citet{Voit15a,Voit15b}. A more direct examination of the thermodynamics of the hot halo gas is to estimate its radiative cooling rate ($\dot{M}_{\rm cool}$). We compute $\dot{M}_{\rm cool}$ of the hot gas within the cooling radius, which are listed in Table~\ref{table:LXRadii}. $\dot{M}_{\rm cool}$ sensitively depends on the slope of the radial intensity profile ($\beta$), which is not well constrained in some cases (e.g., NGC~5908, where a bright nuclear source is removed). However, $\dot{M}_{\rm cool}$ is extremely low ($<1\rm~M_\odot~yr^{-1}$) for all the CGM-MASS galaxies, indicating the radiative cooling in the extended hot gaseous halos cannot be an important gas source to build up the galaxy's stellar content.

\subsection{Missing feedback problem}\label{subsection:MissingFeedback}

Only a small fraction of the SNe feedback energy is detected as X-ray emission in the halo (\S\ref{subsec:EnergyBudget}). We therefore have an apparent ``missing feedback'' problem (e.g., \citealt{Wang10}). There are in general three possible fates of the feedback material: (1) escapes out of the galaxy and joins the intergalactic medium (IGM); (2) cools and falls back to the galactic disk and joins the interstellar medium (ISM); (3) stays in the halo and joins the CGM.

We have shown in \S\ref{subsection:GravitationalHeatingGas} that the gravitational potential of a galaxy as massive as the CGM-MASS galaxies is deep enough so the feedback material typically cannot escape out of the halo. On the other hand, we also show in \S\ref{subsection:Thermodynamics} that the radiative cooling of the halo hot gas is inefficient so the precipitation rate is extremely low for the CGM-MASS galaxies. Therefore, the only possible fate of the feedback material in the CGM-MASS galaxies is to stay in the halo and become part of the CGM.

\begin{figure}
\begin{center}
\epsfig{figure=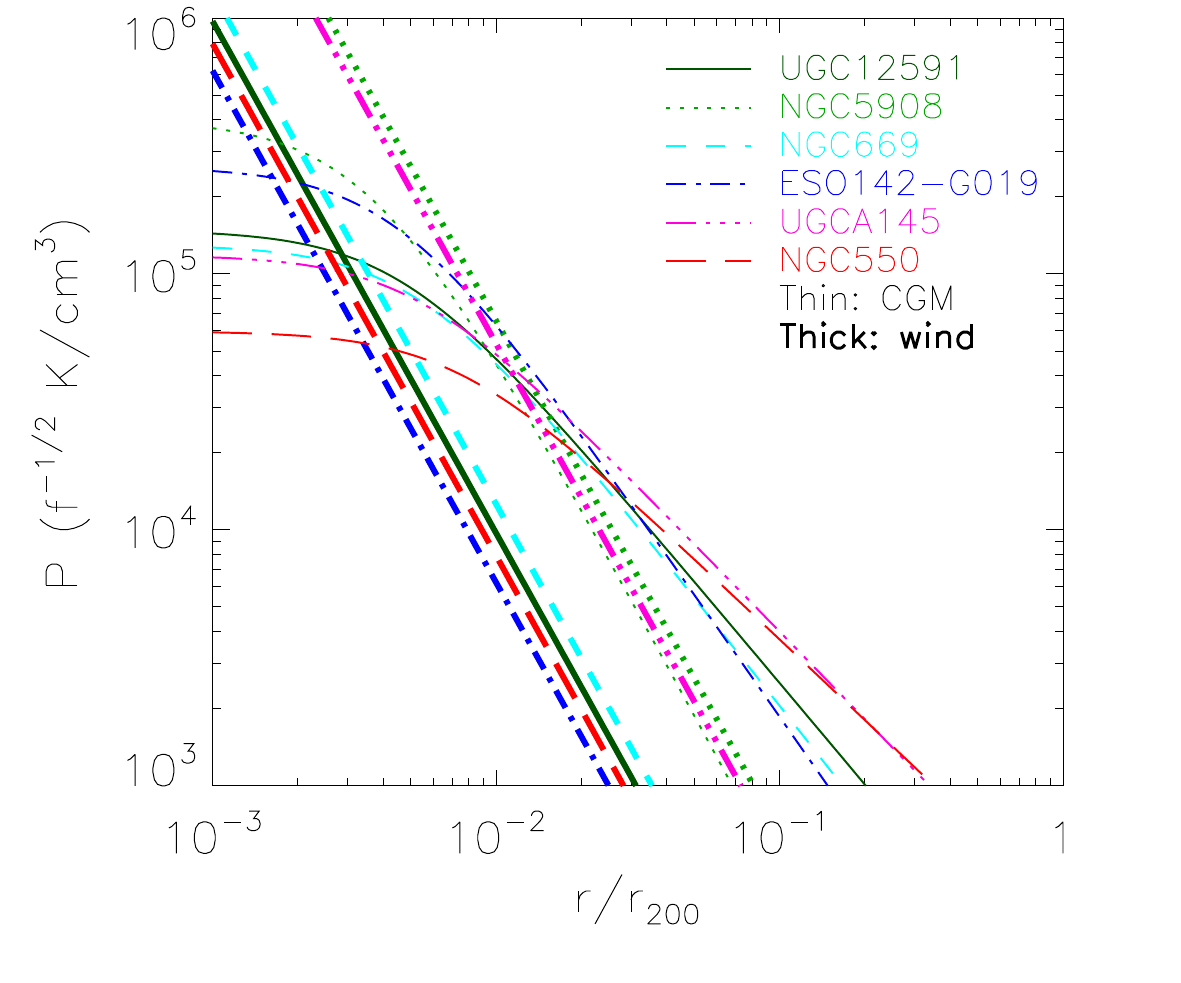,width=0.5\textwidth,angle=0, clip=}
\caption{Hot gas thermal pressure profile (thin curves) calculated with Equ.~\ref{equi:Pbeta} and parameters listed in Tables~\ref{table:SpecPara} and \ref{table:LXRadii}. We also include the thermal pressure of a star formation driven wind blown bubble for comparison (thick curves; \citealt{Veilleux05}).}\label{fig:Pprofile}
\end{center}
\end{figure}

We can further examine the dynamical state of the feedback material by comparing the radial distribution of their thermal pressure to the thermal pressure profile of the pre-existing halo gas. The thermal pressure of the feedback material strongly depends on the star formation properties of the galaxies. We herein use a simple expression of it based on a wind blown bubble scenario, and assume all the star formation happens within the bubble \citep{Veilleux05}. The thermal pressure of the wind and the CGM of the CGM-MASS galaxies are compared in Fig.~\ref{fig:Pprofile}. For most of the CGM-MASS galaxies, $P_{\rm wind}>P_{\rm CGM}$ within a few tens of kpc. However, for NGC~5908, because the density profile is very steep ($\beta\approx0.68$; Table~\ref{table:SpecPara}), the wind may be driven by thermal pressure throughout the halo.

We have assumed a constant temperature when calculating the thermal pressure profile of the CGM, which is certainly oversimplified. The current data does not allow for a constraint on the temperature variation because of the weak hot gas emission and low counting statistic at large radii. \citet{Anderson16}, however, has found a significant radial declination of hot gas temperature in NGC~1961, with $kT$ at $r\approx50\rm~kpc$ about half of the value at $r\approx15\rm~kpc$. If this is also true in the CGM-MASS galaxies, the thermal pressure of the CGM will decline faster at larger radii. Nevertheless, such a temperature drop will at most cause a thermal pressure drop to about half of the value shown in Fig.~\ref{fig:Pprofile} at $r\sim0.1r_{\rm 200}$, which in most of the cases is not large enough to develop a thermal pressure driven wind in the halo.

In addition to a simple wind blown bubble model, we also compare the thermal pressure profile of the CGM to some numerical simulations of low mass galaxies (so the thermal pressure is mainly contributed by the wind). The thermal pressure of a starburst driven wind in \citet{Strickland09} is $\sim10^6\rm~K~cm^{-3}$ at $r\approx0.4\rm~kpc$ ($\sim10^{-3}r_{\rm 200}$ for the CGM-MASS galaxies), on average about one order of magnitude higher than the pressure of the ambient medium in the CGM-MASS galaxies. Therefore, the feedback material can at least expand to a few hundred pc driven by thermal pressure. However, it can unlikely be energetic enough to expand to much larger radii. \citet{Tang09} studied the galactic outflow driven by Type~Ia SNe in star formation inactive galactic bulges, which are more similar to the quiescent CGM-MASS galaxies. They obtained a thermal pressure of $\sim10^5\rm~K~cm^{-3}$ at a galactocentric radius of a few hundred pc, and $\sim10^3\rm~K~cm^{-3}$ at $r\lesssim2\rm~kpc$. This kind of feedback is unlikely energetic enough to expand to a radius larger than a few kpc in the CGM-MASS galaxies. Of course there are some other ways to drive galactic outflows (e.g., \citealt{Breitschwerdt91,Murray05,Krumholz12,Heckman17}), but it is very unlikely that the hot feedback material can be carried out to a significant fraction of the virial radius of such massive galaxies. This is also consistent with a hydrostatic halo as claimed in \S\ref{subsection:GravitationalHeatingGas}. 

We next speculate a scenario to explain the extremely low ($\sim1\%$) X-ray radiation efficiency and the significantly higher X-ray radiation efficiency of NGC~1961 and NGC~6753 than the CGM-MASS galaxies (\S\ref{subsec:EnergyBudget}).
As most of the SNe feedback energy is not dissipated via X-ray radiation, and the feedback material finally mixes with the hot CGM, the X-ray emission is expected to have a large scatter and determined by the density profile instead of the total feedback energy. For a hydrostatic halo with no external gas sources, the SFR is directly linked to the accretion rate of the condensed cool gas. Since the cooling and condensation of the gas produce X-ray emissions, we expect the X-ray radiation efficiency increases with increasing SFR. This is basically different from lower mass galaxies in which the star formation feedback plays a negative role in the X-ray emission of the halo gas, in the way of heating the gas and driving low emissivity galactic superwind \citep{Li13b,Wang16}. There are too few massive spiral galaxies which can host a hydrostatic gaseous halo for a statistical comparison, but the higher X-ray emissivity of NGC~1961 and NGC~6753 than the extremely quiescent CGM-MASS galaxies is apparently consistent with this scenario.

\section{Summary and Conclusions}\label{section:Summary}

We have analyzed the \emph{XMM-Newton} observations of the CGM-MASS galaxies, following the same procedure as presented in Paper~I. We have statistically compared the results of this analysis with those obtained for other galaxy samples to understand the properties of the hot CGM in massive galactic halos. Our main results and conclusions are summarized below. 

$\bullet$ The CGM-MASS galaxies and the MW are consistent with lower mass disk galaxies on the X-ray scaling relations. The $L_{\rm X}$-galaxy mass ($M_*$ or $M_{\rm 200}$) relations of disk galaxies have sublinear slopes, smaller than the slopes of similar relations for elliptical galaxies. The specific X-ray luminosities ($L_{\rm X}/M_*$ or $L_{\rm X}/M_{\rm 200}$) positively correlate with the SFR for most of the disk galaxies, with massive spiral galaxies (CGM-MASS, MW, NGC~1961, NGC~6753) marginally follow the same trend but are always the least X-ray luminous at a given SFR. Study of the scaling relations of the hot gas properties in the outer halo is limited by the number of galaxies with a firm detection of the hot CGM at large radii, but the current result does not conflict with a positive correlation between $L_{\rm X}$ and galaxy mass. Similar as lower mass disk galaxies, typically $\lesssim1\%$ of SNe energy has been converted to soft X-ray emission of the hot gas around quiescent massive spiral galaxies (CGM-MASS, MW), but the X-ray radiation efficiency increases to $\sim5\%$ for the star forming massive spiral NGC~1961 and NGC~6753. 

$\bullet$ The radial distribution of the X-ray emission from hot gas around the CGM-MASS galaxies, after subtracting various stellar and background components, can be well characterized with a $\beta$-function. The radial extension of the hot CGM is typically $\sim(30-100)\rm~kpc$ for individual galaxies above the 1~$\sigma$ background scatter. The CGM-MASS sample thus at least doubles the existing detection of extended hot CGM around massive spiral galaxies. The slope of the radial intensity profile is typically $\beta=0.35-0.55$, except for the slightly higher value of NGC~5908 which is largely affected by the removal of the X-ray bright AGN. $\beta$ of massive spirals (CGM-MASS, MW, NGC~1961, NGC~6753) are all consistent with each other on the plot between $\beta$ and other galaxy properties, and are not significantly different from X-ray luminous elliptical galaxies. The measured $\beta$ of the radial intensity profile of the CGM-MASS galaxies is consistent with those predicted from a hydrostatic isothermal gaseous halo. 

$\bullet$ The diffuse X-ray spectra of the CGM-MASS galaxies at $r<1^\prime-2^\prime$ can be fitted with a thermal plasma model, after removing various fixed stellar and background components. The metallicity of hot gas is poorly constrained and is fixed at $0.2\rm~Z_\odot$ throughout this paper. The temperature of the hot gas is typically $kT\sim0.7\rm~keV$, in the range of $(0.4-1.1)\rm~keV$. $kT$ of low mass disk galaxies is systematically higher than the virial temperature of the host dark matter halo, but massive spirals (CGM-MASS, MW, NGC~1961, NGC~6753) have hot gas temperature comparable to the virial temperature. 

$\bullet$ What is the origin of the halos? A rough estimate indicates that the outflow driven by the thermal pressure of SNe in the CGM-MASS galaxies cannot escape out of the dark matter halo. On the other hand, the ratio between the radiative cooling timescale and the free fall timescale of a cold gas cloud condensed from the hot CGM is much larger than the critical value of $\sim10$ at which the hot CGM can cool and precipitate. Therefore, the hot CGM can neither escape out of the halo nor fall back into the disk. It is mostly likely that the feedback material mixes with the CGM and are both heated gravitationally, forming a hydrostatic galactic corona. The X-ray luminosity of the halo is not directly related to the feedback rate, so there is a large scatter of the X-ray radiation efficiency, which is expected to be positively correlated with the cooling rate so the SFR, but such a trend is not well constrained with the current data.

\bigskip
\noindent\textbf{\uppercase{acknowledgments}}
\smallskip\\
\noindent JTL acknowledges the financial support from NASA through the grants NNX13AE87G, NNH14ZDA001N, and NNX15AM93G. QDW is supported by NASA via a subcontract of the grant NNX15AM93G. RAC is a Royal Society University Research Fellow.

\begin{appendices}

\section{Additional information on XMM-Newton Data Reduction}\label{Appsection:DataReduction}

Details of the \emph{XMM-Newton} data reduction procedures are presented in the appendix of Paper~I. We herein adopt similar data reduction procedures for all the CGM-MASS galaxies. Informations of the \emph{XMM-Newton} data used in this paper are summarized in Table~\ref{table:sampleXMM}.

\begin{table}
\begin{center}
\caption{\emph{XMM-Newton} Data of the CGM-MASS Galaxies.} 
\footnotesize
\vspace{-0.0in}
\begin{tabular}{lcccccccccccccc}
\hline\hline
Galaxy        & ObsID & Start Date & $t_{\rm XMM}$ & $t_{\rm eff,M1}$ & $t_{\rm eff,M2}$ & $t_{\rm eff,PN}$ \\
            &  &  & ks & ks & ks & ks \\
\hline\\
UGC 12591 & 0553870101 & 2008-12-15 & 79.8 & 51.0 & 54.2 & 36.3 \\
NGC 669     & 0741300201 & 2015-02-14 & 123.9 & 83.8 & 92.8 & 54.1 \\
ESO142-G019 & 0741300301 & 2014-09-16 & 91.9 & 73.3 & 75.0 & 55.8 \\
NGC 5908   & 0741300101 & 2014-08-16 & 45.5 & 41.3 & 42.0 & 33.1 \\
UGCA 145   & 0741300401 & 2014-05-21 & 111.6 & 84.4 & 84.6 & 54.7 \\
NGC 550     & 0741300501 & 2015-06-25 & 73.0 & 47.8 & 42.5 & 15.3 \\
                    & 0741300601 & 2015-06-27 & 75.0 & 61.9 & 61.8 & 34.2 \\
\hline\hline
\end{tabular}\label{table:sampleXMM}
\end{center}
$t_{\rm XMM}$ is the total exposures of the \emph{XMM-Newton} observations, while $t_{\rm eff,M1}$, $t_{\rm eff,M2}$, and $t_{\rm eff,PN}$ are the cleaned effective exposure times of MOS-1, MOS-2, and PN respectively.\\
\end{table}

\subsection{Adding the SWCX component in background analysis of some observations}\label{subsection:SWCX}

Background spectra of each galaxy are extracted from the entire FOV, after removing X-ray bright point-like or extended sources. We herein adopt a similar background analysis procedure as in Paper~I for NGC~5908 (see the appendix of Paper~I).

\begin{figure*}
\begin{center}
\hspace{-0.3in}
\epsfig{figure=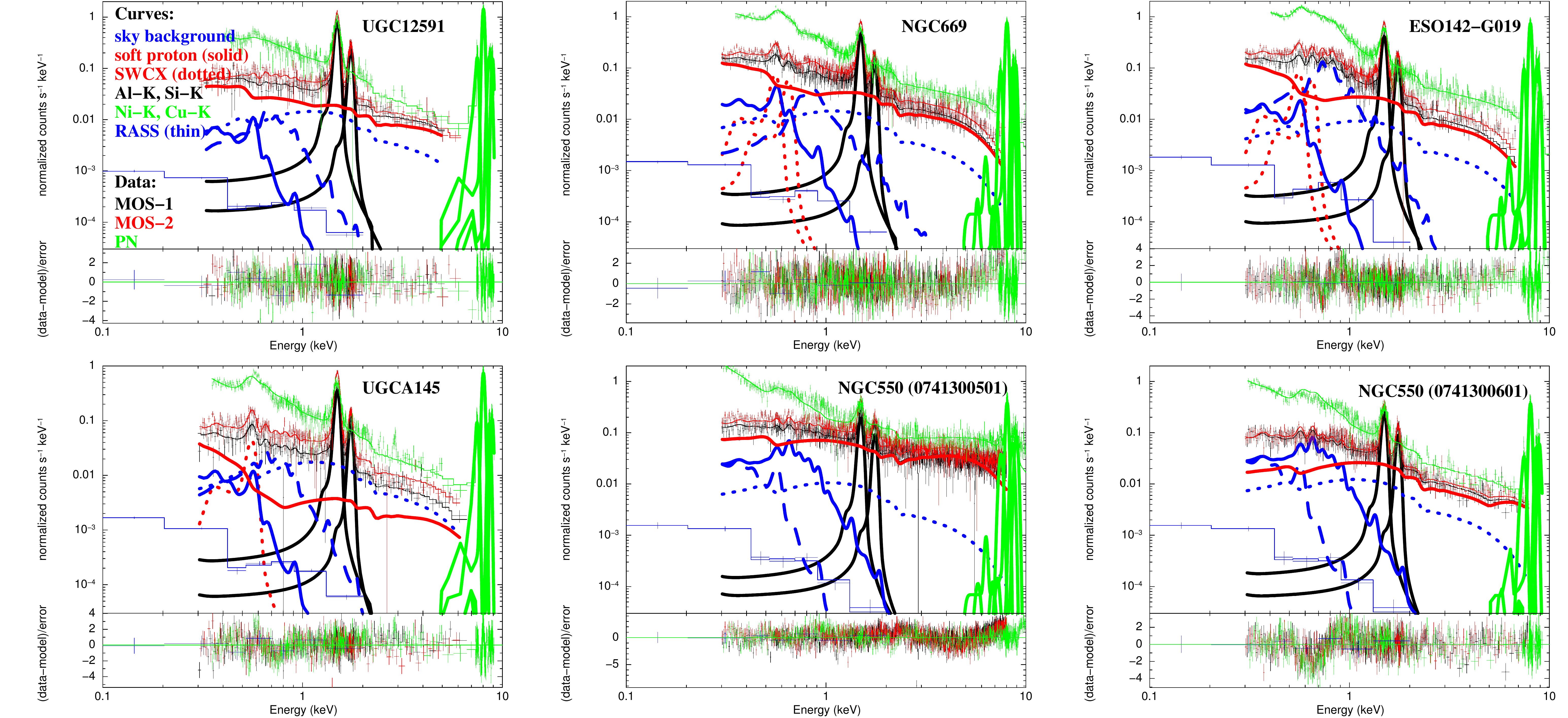,width=1.0\textwidth,angle=0, clip=}
\caption{Background spectra of the CGM-MASS galaxies extracted from a large enough annulus to exclude the emission associated with the galaxy. We also remove point sources and prominent diffuse X-ray features using the masks shown in Fig.~\ref{fig:SourceDiffuseMask}. These spectra are similar as the background spectra of NGC~5908 presented in Paper~I, but we add a SWCX component for some galaxies. The \emph{ROSAT} all sky survey (RASS) spectrum extracted from a $0.2^\circ-1^\circ$ annulus centered at the galaxy is also included in order to help constraining the sky background (blue data points and thin solid curve). Curves of different model components are scaled to the MOS-1 spectrum, except for the Ni and Cu~K$\alpha$ lines, which are scaled to the PN spectrum. Colored data points with error bars are spectra of MOS-1 (black), MOS-2 (red), and PN (green), respectively. Colored curves denote different background model components: sky background including the local hot bubble (blue solid), the Galactic halo (blue dashed), and the distant AGN (blue dotted) components, soft proton (red), SWCX (red dotted), Al-K$\alpha$ and Si-K$\alpha$ instrumental lines (two black gaussian lines), and Ni-K$\alpha$ and Cu-K$\alpha$ lines of PN only (four green gaussian lines). The two observations of NGC~550 are plotted in separated panels.}\label{fig:bckspec}
\end{center}
\end{figure*}

The \emph{XMM-Newton} data of NGC~5908 does not have a significant solar wind charge exchange (SWCX) component in the background spectra. This component, however, is important for the \emph{XMM-Newton} observations of some of our sample galaxies (ESO142-G019, NGC~669, and UGCA~145). We add two gaussian lines with zero line width at 0.56~keV and 0.65~keV to represent the SWCX contribution, following the \emph{XMM-Newton} background analysis cookbook (\url{ftp://xmm.esac.esa.int/pub/xmm-esas/xmm-esas.pdf}). The fitted background spectra are shown in Fig.~\ref{fig:bckspec}, with all other components (distant AGN, MW halo, local hot bubble, soft proton, and instrumental lines) the same as described in Paper~I. Such a background analysis is not aiming at physically decomposing and modeling various background components in the most accurate way, but at roughly characterizing the background in an identical way for different galaxies in order to quantitatively subtract different background components in spatial and spectral analysis.

We further create a SWCX image following the steps described in the ESAS data analysis threads (\url{http://www.cosmos.esa.int/web/xmm-newton/sas-thread-esasimage}). The best-fit normalizations of the gaussian lines are rescaled with the area of the background spectral extraction regions, with prominent point-like or extended X-ray features removed. We then create the SWCX image with the {\small SAS} task {\small swcx}. This SWCX image is adopted in the follow-up imaging (\S\ref{subsection:XMMImages}) and spatial analyses (\S\ref{subsection:SpatialCorona}).

\subsection{Prominent X-ray features of the sample galaxies}\label{subsection:XMMImages}

\begin{figure*}
\begin{center}
\epsfig{figure=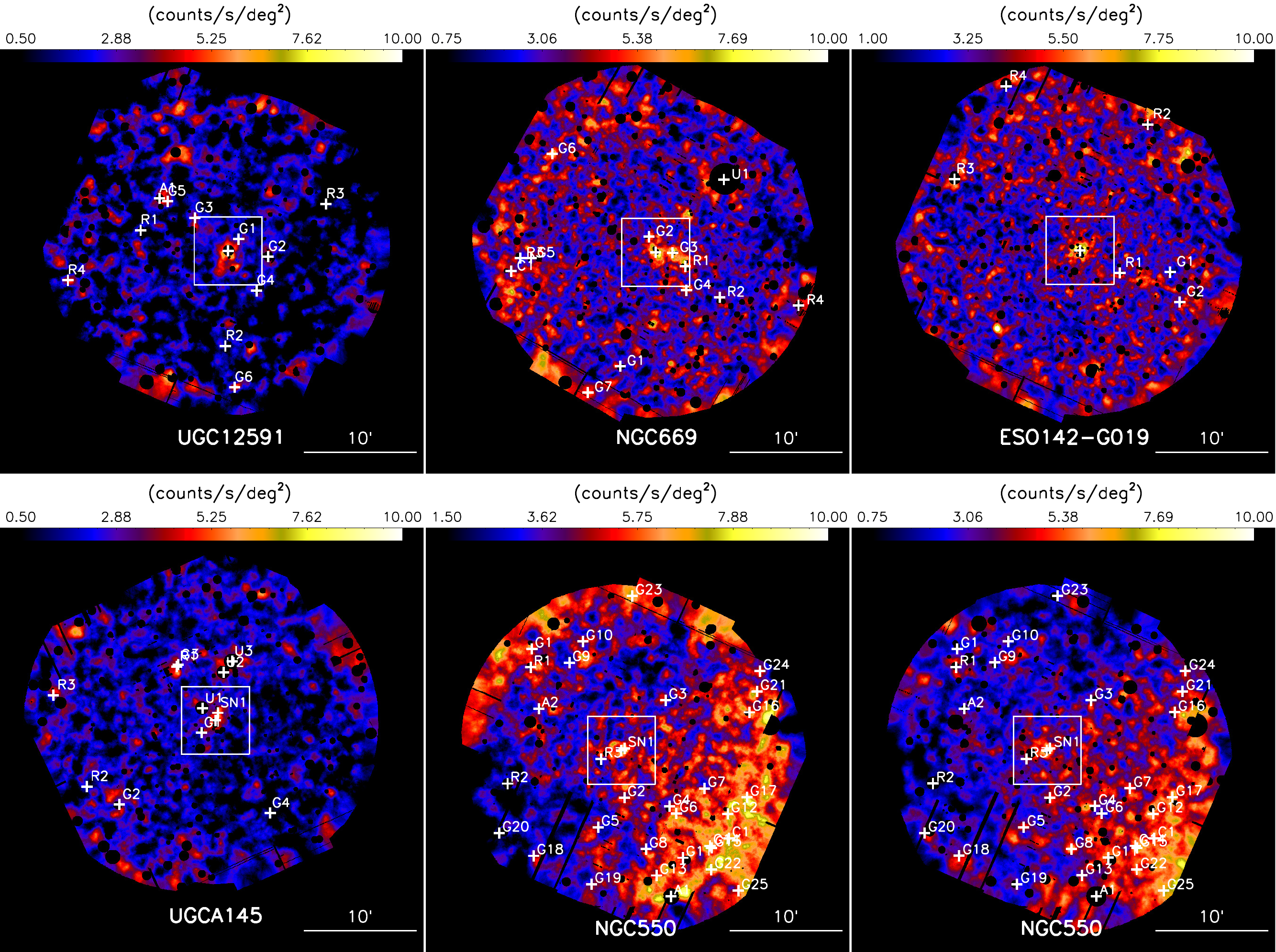,width=1.0\textwidth,angle=0, clip=}
\caption{Point source removed, soft proton and quiescent particle background subtracted, exposure corrected, and adaptively smoothed 0.5-1.25~keV \emph{XMM-Newton} EPIC (MOS-1+MOS-2+PN) image of the CGM-MASS galaxies and the surrounding area (a similar image of NGC~5908 is presented in Paper~I). The color bar in $\rm counts~s^{-1}~deg^{-2}$ is marked on top of the images. The exposure maps of different instruments are scaled to MOS-2 with the medium filter before the mosaicing. Cross identified sources are marked with pluses, which are also listed in Table~\ref{table:IdentifiedSources}. The white box in each panel marks the central $6^\prime\times6^\prime$ region, the close-up of which is shown in Fig.~\ref{fig:imagesZoomin}. The last two panels are for different observations of NGC~550, which are separated by two days (ObsID 0741300501 and 0741300601; Table~\ref{table:sampleXMM}).}\label{fig:imagesFOV}
\end{center}
\end{figure*}

We present the soft proton and quiescent particle background subtracted, exposure corrected, and adaptively smoothed 0.5-1.25~keV \emph{XMM-Newton} images of the entire FOV of individual observations in Fig.~\ref{fig:imagesFOV}. The images are primarily used to show the environment of the galaxies and the cross-identified foreground or background sources. A zoom-in of the apparently diffuse X-ray emission largely from hot gas associated with the target galaxies are presented in Fig.~\ref{fig:imagesZoomin}. Detected point-like X-ray sources are removed with circular masks in Fig.~\ref{fig:imagesFOV}. The brightest point sources are summarized in Table~\ref{table:IdentifiedSources}, while the properties of all the detected X-ray sources are available in the online catalog. Below we briefly describe the most important ones of them.

\textbf{UGC~12591:} The diffuse soft X-ray emission around this galaxy is slightly elongated along its minor axis, especially on the southeast side (Fig.~\ref{fig:imagesZoomin}). However, this elongation may be largely contaminated by the residual emission of a removed X-ray bright foreground star. We therefore extract the spectra of the hot halo only from a circle with $r<1^\prime$ (Fig.~\ref{fig:imagesZoomin}). 

\begin{table*}[t]{}
\begin{center}
\caption{Cross Identified Sources Around the CGM-MASS Galaxies} 
\label{table:IdentifiedSources}
\footnotesize
\tabcolsep=3.pt%
\begin{tabular}{lllcccccccccccc}
\hline \hline
Galaxy        & No.     & Identified Source Name & RA,DEC (J2000.0) & Type & Redshift & X-ray Shape \\
$L_{\rm 0.3-7.2keV,limit}$        &     &    &  &  &  &  \\
$10^{38}\rm~ergs~s^{-1}$        &     &    &  &  &  &  \\\hline
UGC 12591   & G1 & 2MASX J23251750+2830445 & 23:25:17.5,+28:30:44 & Galaxy    & -              & Point \\
5.86               & G2 & 2MASX J23250574+2829115 & 23:25:05.7,+28:29:12 & Galaxy    & -              & Point? \\
		    & G3 & 2MASX J23253520+2832368 & 23:25:35.2,+28:32:37 & Galaxy    & -              & Extended? \\
		    & G4 & 2MASX J23251029+2826105 & 23:25:10.3,+28:26:11 & Galaxy    & -              & Point \\
		    & G5 & 2MASX J23254604+2834048 & 23:25:46.1,+28:34:05 & Galaxy    & -              & Extended \\
		    & G6 & 2MASX J23251916+2817375 & 23:25:19.1,+28:17:38 & Galaxy    & -              & Extended \\
		    & A1 & 2MASX J23254938+2834208 & 23:25:49.4,+28:34:21 & Seyfert 2 & 0.114006 & Point+Extended \\
		    & R1 & NVSS J232557+283131         & 23:25:57.0,+28:31:32 & Radio Source    & -               & Point \\
		    & R2 & AGC 333535			  & 23:25:22.9,+28:21:17 & Radio Source    & 0.024067 & Offset Point \\
		    & R3 & NVSS J232442+283350         & 23:24:42.4,+28:33:50 & Radio Source    & -               & Point \\
		    & R4 & B2 2323+28			  & 23:26:26.2,+28:27:05 & Radio Source    & -               & Extended \\
\hline
NGC 669      & G1 & 2MASX J01473153+3523428 & 01:47:31.5,+35:23:43 & Galaxy    & -               & Point \\
2.48		   & G2 & [WGB2006] 014400+34320e   & 01:47:19.1,+35:35:11 & Galaxy    & 0.10114    & Point+Extended \\
		   & G3 & 2MASX J01470878+3533448 & 01:47:08.8,+35:33:45 & Galaxy    & -                & Point+Extended \\
		   & G4 & 2MASX J01470281+3530268 & 01:47:02.8,+35:30:26 & Galaxy    & -                & Point+Extended \\
		   & G5 & 2MASX J01481042+3533165 & 01:48:10.4,+35:33:17 & Galaxy    & -                & Point \\
		   & G6 & KUG 0145+354                       & 01:48:01.3,+35:42:30 & Galaxy    & -                & Extended \\
		   & G7 & 2MASX J01474564+3521239 & 01:47:45.6,+35:21:23 & Galaxy    & -                & Extended? \\
		   & C1 & PPS2 118				  & 01:48:19.1,+35:32:07 & Galaxy Group & 0.013993 & Offset Point \\
		   & R1 & NVSS J014703+353232          & 01:47:03.3,+35:32:33 & Radio Source   & -                & Multi Point+Extended \\
		   & R2 & NVSS J014648+352948          & 01:46:48.4,+35:29:48 & Radio Source   & -                & Multi Offset Point \\
		   & R3 & NVSS J014815+353316	  & 01:48:15.1,+35:33:16 & Radio Source   & -                & Point \\
		   & R4 & NVSS J014614+352905	  & 01:46:14.1,+35:29:05 & Radio Source   & -                & Point \\
		   & U1 & -                                               & 01:46:46.5,+35:40:13 & Unidentified & -           & Point \\
\hline
ESO142-G019 & G1 & 2MASX J19321841-5808437 & 19:32:18.4,-58:08:44 & Galaxy    & -                & Point \\
1.83		       & G2 & 2MASX J19321186-5811238 & 19:32:11.9,-58:11:23 & Galaxy    & -                & Point \\
		       & R1 & SUMSS J193252-580849      & 19:32:52.1,-58:08:50 & Radio Source   & -                & Point \\
		       & R2 & SUMSS J193233-575540      & 19:32:33.8,-57:55:41 & Radio Source   & -                & Point \\
		       & R3 & PMN J1934-5800                   & 19:34:42.8,-58:00:31 & Radio Source   & -                & Point? \\
		       & R4 & SUMSS J193408-575218      & 19:34:08.1,-57:52:18 & Radio Source   & -                & Point \\
\hline
UGCA 145    & G1 & LEDA 846894                         & 08:47:22.2,-20:03:13 & Galaxy    & -                & Point? \\
2.11		    & G2 & 2MASX J08475323-2009342 & 08:47:53.2,-20:09:34 & Galaxy    & 0.04073    & Point+Extended? \\
		    & G3 & ESO 563-22                           & 08:47:31.0,-19:57:12 & Galaxy    & -                & Point+Extended \\
		    & G4 & 2MASX J08465630-2010192 & 08:46:56.3,-20:10:20 & Galaxy    & -                & Point \\
		    & R1 & NVSS J084731-195722         & 08:47:31.5,-19:57:22 & Radio Source & -         & Point \\
		    & R2 & NVSS J084805-200758	 & 08:48:05.4,-20:07:59 & Radio Source & -         & Point \\
		    & R3 & NVSS J084818-195954	 & 08:48:18.0,-19:59:54 & Radio Source & -         & Offset Point+Extended? \\
		    & U1 & -                                              & 08:47:21.8,-20:01:04 & Unidentified & -           & Point+Extended \\
		    & U2 & -                                              & 08:47:13.9,-19:57:53 & Unidentified & -           & Point+Extended \\
		    & U3 & -                                              & 08:47:10.7,-19:56:54 & Unidentified & -           & Point+Extended \\
		   & SN1 & SN 2007sq (in UGCA 145)   & 08:47:16.1,-20:01:28 & Supernova  & 0.015274 & Offset Point+Extended? \\
\hline
NGC 550     & G1 & 2MASX J01271427+0210193 & 01:27:14.3,+02:10:19 & Galaxy    & 0.045037  & Offset Point?+Extended? \\
3.27		  & G2 & GALEXASC J012641.58+015709.8 & 01:26:41.5,+01:57:10 & Galaxy & -          & Point?+Extended? \\
		  & G3 & [HC2009] 01950                      & 01:26:26.9,+02:05:47 & Galaxy Pair & -            & Point \\
		  & G4 & GALEXASC J012625.59+015625.1 & 01:26:25.5,+01:56:25 & Galaxy & -          & Offset Point \\
		  & G5 & APMUKS(BJ) B012416.29+013900.2 & 01:26:50.8,+01:54:33 & Galaxy & -       & Extended \\
		  & G6 & 2MASX J01262301+0155465 & 01:26:23.1,+01:55:46 & Galaxy     & 0.174000 & Point+Extended \\
		  & G7 & GALEXASC J012613.12+015759.3 & 01:26:13.1,+01:57:59 & Galaxy & -          & Extended? \\
		  & G8 & 2MASX J01263382+0152375 & 01:26:33.8,+01:52:37 & Galaxy     & -               & Point?+Extended? \\
		  & G9 & GALEXASC J012701.01+020908.3 & 01:27:01.0,+02:09:07 & Galaxy & -          & Point? \\
		  & G10 & GALEXASC J012656.32+021100.4 & 01:26:56.2,+02:10:59 & Galaxy Pair & - & Point? \\
		  & G11 & 2MASX J01262088+0151505 & 01:26:20.8,+01:51:51 & Galaxy       & -           & Point+Extended \\
		  & G12 & GALEXASC J012604.94+015544.6 & 01:26:04.9,+01:55:44 & Galaxy & -        & Offset Point?+Extended \\
		  & G13 & [BDG98] J012630.1+015017  & 01:26:30.1,+01:50:17 & Galaxy    & 0.018389 & Extended? \\
		  & G14 & GALEXASC J012611.24+015253.6 & 01:26:11.2,+01:52:54 & Galaxy & -         & Extended \\
		  & G15 & 2MASX J01261063+0152430 & 01:26:10.6,+01:52:43 & Galaxy    & -               & Extended \\
		  & G16 & GALEXASC J012557.30+020444.8 & 01:25:57.3,+02:04:43 & Galaxy & -        & Offset Point+Extended? \\
		  & G17 & GALEX 2673389968112487937 & 01:25:58.1,+01:57:12 & Galaxy      & -         & Extended \\
		  & G18 & GALEXASC J012713.58+015202.0 & 01:27:13.6,+01:52:01 & Galaxy & -        & Extended? \\
		  & G19 & GALEXASC J012653.12+014928.4 & 01:26:53.1,+01:49:30 & Galaxy & -        & Offset Point? \\
		  & G20 & GALEXASC J012725.75+015403.1 & 01:27:25.8,+01:54:02 & Galaxy & -        & Offset Point \\
		  & G21 & GALEXASC J012554.54+020636.4 & 01:25:54.5,+02:06:34 & Galaxy & 0.006131 & Extended \\
		  & G22 & APMUKS(BJ) B012336.28+013514.2 & 01:26:10.7,+01:50:48 & Galaxy & -      & Extended \\
		  & G23 & [HC2009] 01970                      & 01:26:38.7,+02:15:01 & Galaxy Pair & -          & Point?+Extended \\
		  & G24 & GALEXASC J012553.56+020822.5 & 01:25:53.5,+02:08:22 & Galaxy & -         & Extended? \\
		  & G25 & APMUKS(BJ) B012326.79+013321.4 & 01:26:01.2,+01:48:56 & Galaxy & -      & Extended \\
		  & C1 & [M98j] 019                                 & 01:26:04.5,+01:53:34 & Galaxy Group & 0.018339 & Extended \\
		  & A1 & 2XMM J012624.9+014825        & 01:26:25.0,+01:48:26 & AGN Candidate & -         & Point \\
		  & A2 & SDSS J012711.79+020501.6    & 01:27:11.8,+02:05:02 & QSO		    & 2.757000 & Point \\
		  & R1 & PMN J0127+0208                     & 01:27:14.7,+02:08:42 & Radio Source & -        & Point+Extended? \\
		  & R2 & PMN J0127+0158			  & 01:27:22.9,+01:58:25 & Radio Source & -        & Point \\
		  & R3 & NVSS J012649+020035           & 01:26:49.8,+02:00:35 & Radio Source & -        & Point \\
		  & SN1 & SN 1961Q (in NGC 550)        & 01:26:41.4,+02:01:32 & Supernova      & 0.019443 & Point \\
\hline \hline
\end{tabular}
\end{center}
\vspace{-0.1in}
$L_{\rm 0.3-7.2keV,limit}$ is the point source detection limit in 0.3-7.2~keV, assuming 10~counts per source for a firm detection. Locations of the sources are shown in Fig.~\ref{fig:imagesFOV}, with their No. marked beside. The sources are all identified in NED. We classify the X-ray shape of the source as Point or Extended, and mark those with significant uncertainties on this classification with ``?''. Sometimes there are more than one X-ray source close to, or the X-ray source is offset to the identified source. In these cases, we add ``Multi'' or ``Offset'' to their classifications. 
\end{table*}

\textbf{NGC~669:} Diffuse X-ray emission apparently associated with NGC~669 is significantly elongated toward west (Fig.~\ref{fig:imagesZoomin}). There are two X-ray bright point-like sources probably associated with this extended feature, one of which is an identified galaxy (G3), while the other one is a radio source (R1 in Fig.~\ref{fig:imagesFOV}; Table~\ref{table:IdentifiedSources}). Although we do not have distance estimates of these sources, it is very likely that the extended feature is not related to NGC~669. Therefore, we extract the X-ray spectra of NGC~669 from an elliptical region excluding the feature (Fig.~\ref{fig:imagesZoomin}). It is also excluded in later spatial analysis (\S\ref{subsection:SpatialCorona}; Fig.~\ref{fig:SourceDiffuseMask}). Discussion on the unidentified X-ray bright point-like source U1 will be presented in \S\ref{subsection:Xpointsrc}.

\textbf{ESO142-G019:} Diffuse X-ray emission features around this galaxy apparently extend to at least $\sim1^\prime$ from the galactic center (Fig.~\ref{fig:imagesZoomin}). The galaxy has relatively low Galactic latitude ($b\approx-28^\circ$), so there are more Galactic foreground stars projected in front of it than other CGM-MASS galaxies. There are a few X-ray bright point sources detected close to the galaxy; at least some of them can be attributed to these Galactic stellar sources. The surrounding area of ESO142-G019 is relatively clean without any significant unrelated diffuse features. We therefore extract spectra from a $r=1^\prime$ circle which enclose the prominent diffuse X-ray emission features (Fig.~\ref{fig:imagesZoomin}).

\textbf{UGCA~145:} There is an X-ray bright point source to the east of UGCA~145 (U1; Fig.~\ref{fig:imagesFOV}; Table~\ref{table:IdentifiedSources}; \S\ref{subsection:Xpointsrc}), which is probably responsible for some strange extended X-ray emission features in this area. There are also two unidentified X-ray bright point sources to the north of the galaxy (U2 and U3; \S\ref{subsection:Xpointsrc}), which may also produce some apparently extended features. We remove these sources and other point-like sources from both spectral and spatial analysis. There is an X-ray counterpart of the supernova remnant SN2007sq detected in the galactic disk.

\textbf{NGC~550:} There is a background massive galaxy cluster Abell~189 ($d=132\rm~Mpc$, $z=0.0328$) with its center projected $\sim0.43^\circ$ southwest to NGC~550. The X-ray emission from the cluster extends significantly to the southwest of NGC~550. We remove this region in our spatial and background analysis. There is still noticeable X-ray enhancement related to NGC~550, which is attributed to the galactic corona in the present paper. The supernova remnant SN1961Q is too close to the galactic center to determine which one of them or both correspond to the X-ray peak at the center of the galaxy. The X-ray emission of this nuclear source does appear to be mostly thermal (see more discussions in \S\ref{subsection:Xpointsrc}).

In summary, thanks to their edge-on orientation, all the CGM-MASS galaxies have extraplanar diffuse X-ray emission detected. Although the edge-on galactic disk may absorb a significant fraction of the soft X-ray emission from stellar X-ray sources, weak point-like X-ray source is seen at the center of each galaxy. 

\begin{figure*}
\begin{center}
\epsfig{figure=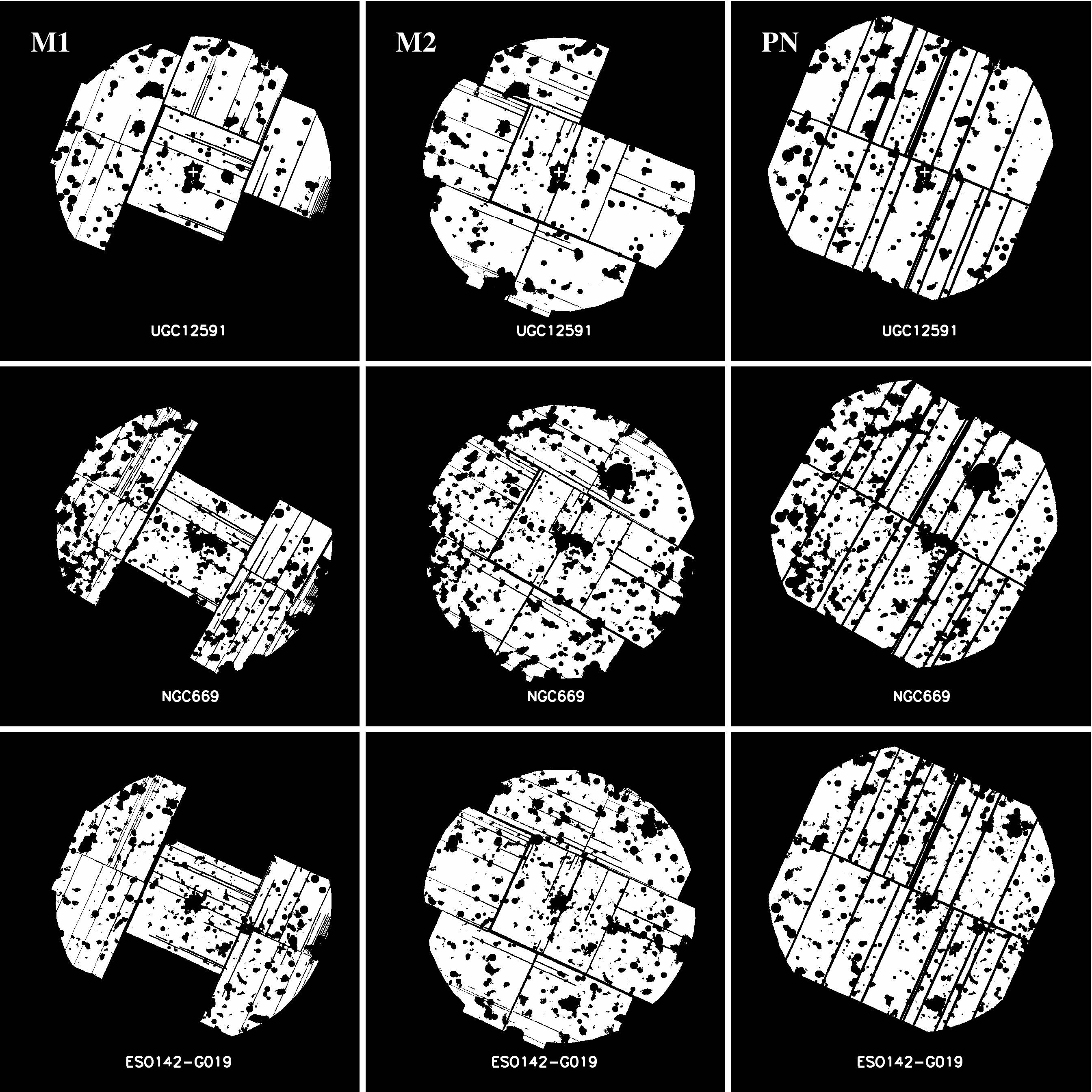,width=1.0\textwidth,angle=0, clip=}
\caption{A combination of point source masks and the masks of prominent diffuse soft X-ray emission features (Fig.~\ref{fig:imagesFOV}). Black regions are filtered out when doing spatial and spectral analyses (\S\ref{subsection:SpatialCorona}, \ref{subsection:SpecCorona}), except for the white circular regions shown in Fig.~\ref{fig:imagesZoomin} which are used to extract halo spectra. The three columns from left to right are for MOS-1, MOS-2, and PN, while different rows show different galaxies. Similar figures of NGC~5908 are presented in the Appendix of Paper~I. NGC~550 has two observations. The ObsID of each observation is denoted in the left column of the related rows.}\label{fig:SourceDiffuseMask}
\end{center}
\end{figure*}
\addtocounter{figure}{-1}
\begin{figure*}
\begin{center}
\epsfig{figure=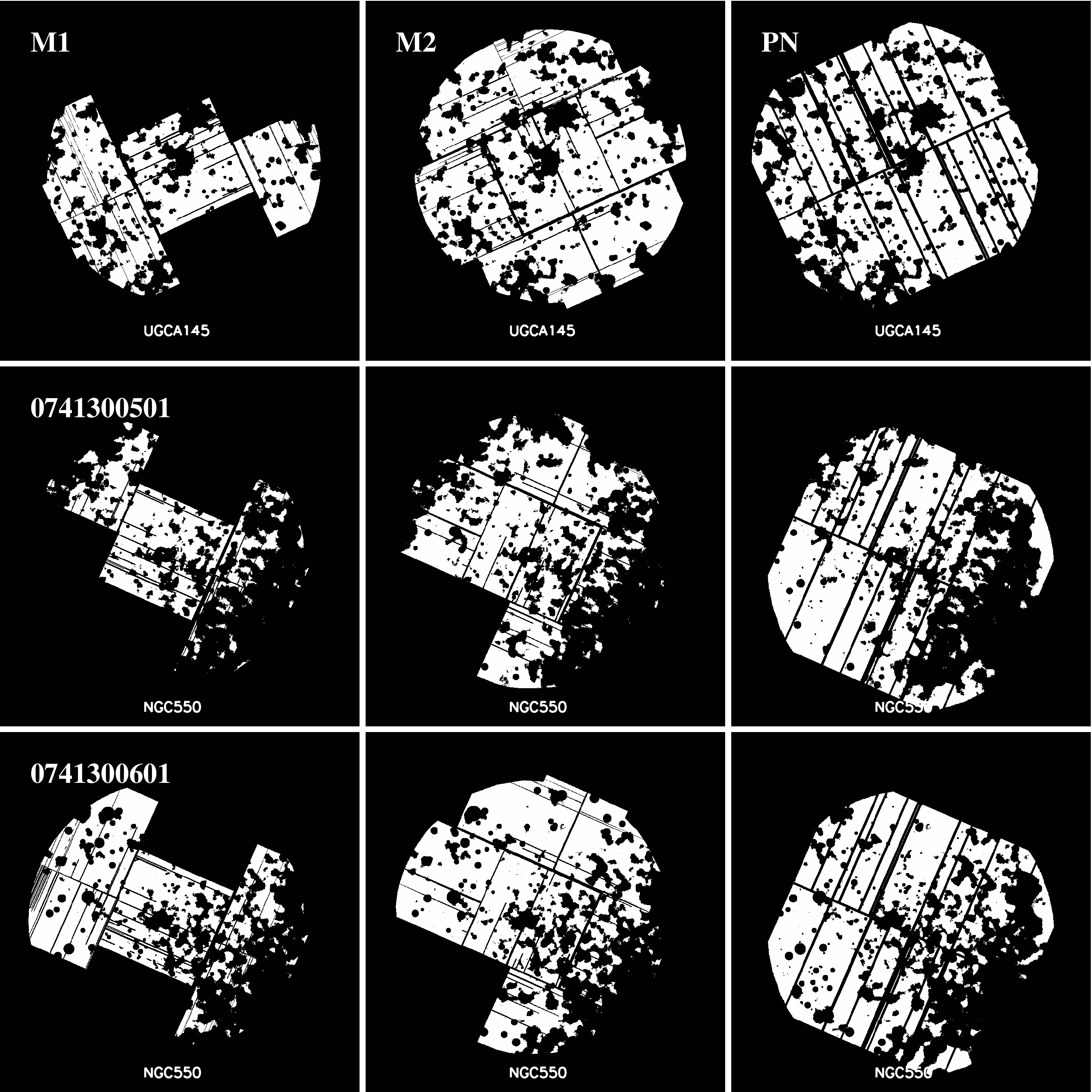,width=1.0\textwidth,angle=0, clip=}
\caption{continued.}
\end{center}
\end{figure*}

\subsection{X-ray bright point-like sources}\label{subsection:Xpointsrc}

There are a few point-like sources which are bright enough in X-ray for us to perform spectral analysis. We present their \emph{XMM-Newton} spectra and discuss the properties of these sources below.

\begin{figure*}
\begin{center}
\epsfig{figure=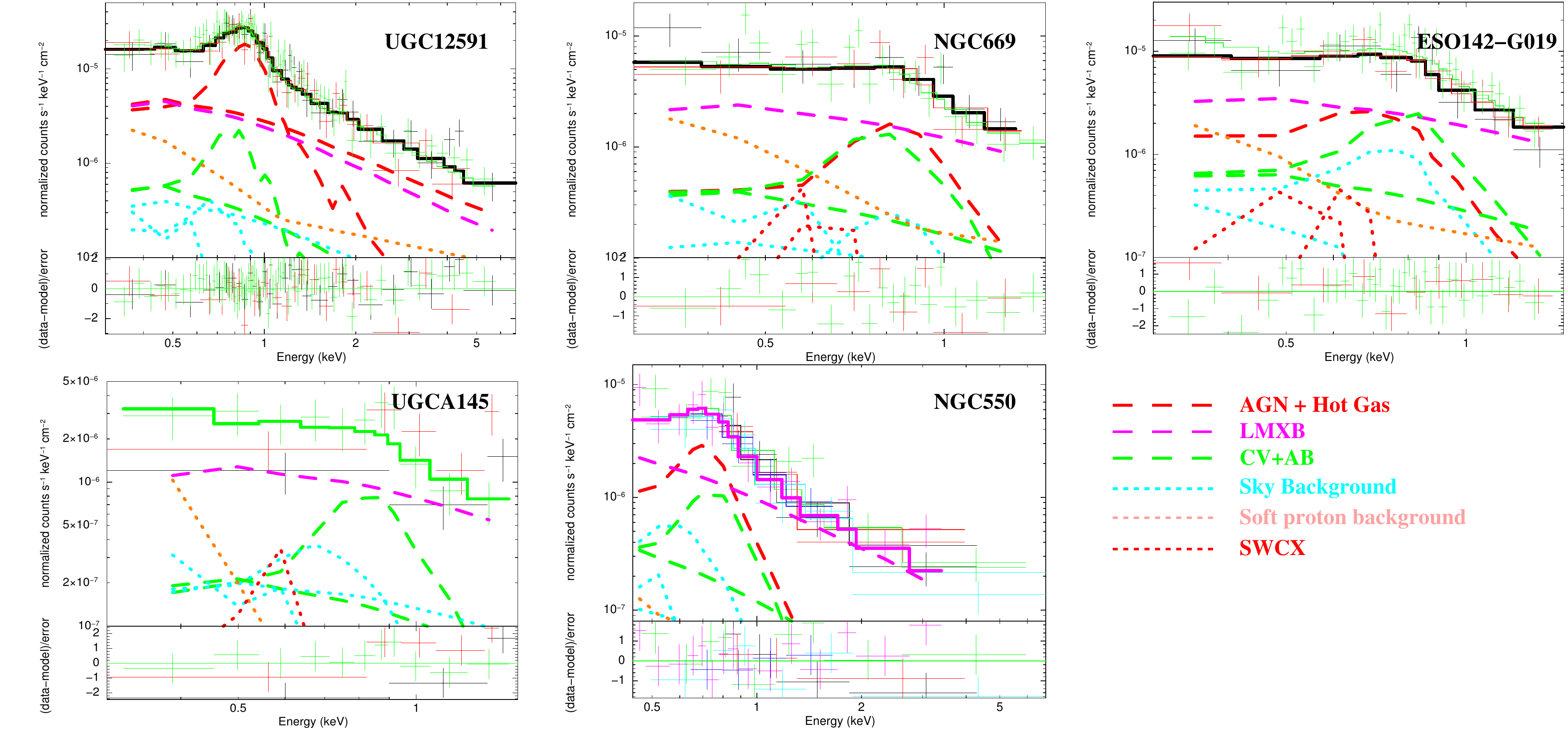,width=1.0\textwidth,angle=0, clip=}
\caption{AGN spectra extracted from the nuclear region of each galaxy. Curves representing different model components are denoted in the lower right panel. Data points and their scalings are the same as in Fig.~\ref{fig:HaloSpec}.}\label{fig:AGNspec}
\end{center}
\end{figure*}

We first present the \emph{XMM-Newton} spectra extracted from the nuclear regions of the CGM-MASS galaxies in Fig.~\ref{fig:AGNspec}. The extraction regions are chosen to have radii of typically $0.2^\prime-0.5^\prime$ around the central X-ray peaks of the galaxies (Fig.~\ref{fig:imagesFOV}). Within the regions, the stellar contribution from the galactic bulges could be significant and can be estimated by scaling the enclosed K-band luminosity, which are discussed in \S\ref{subsection:SpatialCorona}. We find that this contribution is negligible for NGC~5908 (Paper~I), which is the closest one in our sample. We describe the hot gas emission enclosed in each of the regions with an ``APEC'' model \citep{Smith01} and the possible AGN contribution with a power law model. All these components are subjected to the Galactic foreground absorptions. We do not find significant evidence for additional absorption intrinsic to the host galaxies. Only UGC~12591 and NGC~5908 (Paper~I) show significant AGN contributions. The nuclear spectra of UGCA~145 can be reproduced well with the estimated stellar contribution alone, in addition to the fixed background components. For the other galaxies, the spectra need a thermal plasma, representing a putative diffuse hot gas contribution in the bulge regions. The best-fit parameters of the hot gas and AGN in the nuclear region of the CGM-MASS galaxies (except for UGCA~145) are summarized in Table~\ref{table:AGN}. The AGN of NGC~5908 has a strong Fe~K line (Paper~I), with a 6-7~keV luminosity of $8.61_{-0.90}^{+0.91}\times10^{39}\rm~ergs~s^{-1}$, which is not listed in Table~\ref{table:AGN}. We do not detect an Fe~K line in other galaxies (Fig.~\ref{fig:AGNspec}).

\begin{table}
\begin{center}
\caption{X-ray Properties of the Nuclear Regions of the CGM-MASS Galaxies.} 
\footnotesize
\vspace{-0.0in}
\begin{tabular}{lcccccccccccccc}
\hline\hline
Galaxy        & $L_{\rm X,power}$ & $L_{\rm X,APEC}$ & $\Gamma$ & $kT$ \\
            & $10^{39}\rm~ergs~s^{-1}$ & $10^{39}\rm~ergs~s^{-1}$ & & keV \\
\hline
UGC 12591 & $27.7_{-3.0}^{+3.3}$ & $13.9_{-1.2}^{+1.1}$ & $1.44_{-0.18}^{+0.22}$ & $0.82_{-0.03}^{+0.04}$ \\
NGC 669     & - & $0.82\pm0.23$ & - &  $0.77_{-0.28}^{+0.14}$ \\
ESO142-G019 & - & $0.87_{-0.19}^{+0.18}$ & - & $0.30_{-0.04}^{+0.06}$ \\
NGC 5908   & $8.17_{-0.74}^{+0.76}$ & $0.58\pm0.18$ & $1.26\pm0.12$ & $0.81_{-0.13}^{+0.14}$ \\
NGC 550     & - & $1.44_{-0.29}^{+0.27}$ & - & $0.37_{-0.08}^{+0.14}$ \\
\hline\hline
\end{tabular}\label{table:AGN}
\end{center}
The net (background and stellar components subtracted) spectra of the regions are modeled with a power law plus a thermal plasma (APEC), if needed. The key parameters of these two components are represented by $\Gamma$ (photon index) and $kT$ (hot gas temperature), while their corresponding absorption-corrected luminosities are given in the 0.3-8 keV and 0.5-2 keV bands, respectively. 
\end{table}

\begin{figure*}
\begin{center}
\epsfig{figure=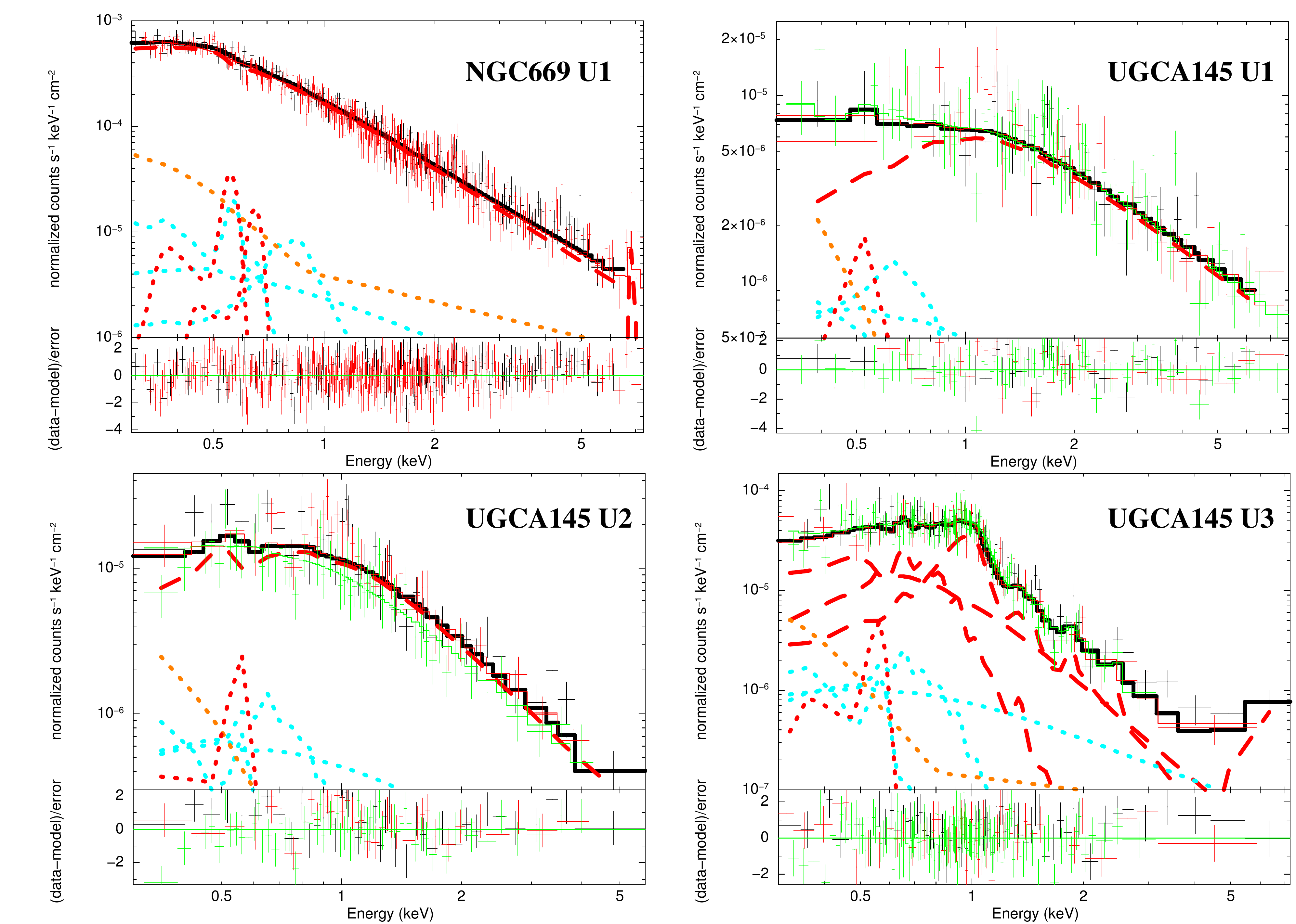,width=0.67\textwidth,angle=0, clip=}
\caption{\emph{XMM-Newton} spectra of X-ray bright unidentified point-like sources. Locations of the sources are shown in Fig.~\ref{fig:imagesFOV} and listed in Table~\ref{table:IdentifiedSources}. Symbols are the same as in Fig.~\ref{fig:AGNspec}. MOS-1 does not cover NGC~669 U1 (due to the removal of noise CCD), so only MOS-2 (black) and PN (red) spectra are shown.}\label{fig:PointSrcSpec}
\end{center}
\end{figure*}

We also extract \emph{XMM-Newton} spectra from the brightest unidentified X-ray sources in the FOV of the observations (Fig.~\ref{fig:imagesFOV}), including Source U1 close to NGC~669, and Sources U1, U2, and U3 close to UGCA~145 (Fig.~\ref{fig:PointSrcSpec}). None of these sources are close enough to the target galaxies to be likely their stellar sources.

Source U1 of NGC~669 can be well fitted with a power law subjected to Galactic foreground absorption plus a gaussian line centered at $6.83\pm0.05\rm~keV$. It has a very faint optical/near-IR counterpart with J, H, and K band magnitudes of 19.5, 16.8, and 16.1. If the source is intrinsically as luminous as the Sun in K-band, the measured K-band apparent magnitude will put it at 3.7~kpc from us, within the MW halo. Therefore, this source is either a foreground stellar source or a background AGN.

\begin{table}
\begin{center}
\caption{X-ray Properties of Bright Unidentified Sources.} 
\footnotesize
\vspace{-0.0in}
\begin{tabular}{lcccccccccccccc}
\hline\hline
Galaxy        & $N_{\rm H}$ & $F_{\rm X,power}$ & $\Gamma$ \\
            & $10^{20}\rm~cm^{-2}$ & $10^{-13}\rm~ergs~s^{-1}~cm^{-2}$ & \\
\hline
NGC 669 U1    & 5.04 (fixed) & $9.44_{-0.12}^{+0.11}$ & $2.24_{-0.02}^{+0.01}$ \\
UGCA 145 U1    & $27.4_{-4.0}^{+3.4}$ & $0.87\pm0.04$ & $1.47_{-0.09}^{+0.06}$ \\
UGCA 145 U2    & $31.0_{-3.4}^{+3.9}$ & $1.18_{-0.14}^{+0.21}$ & $2.82_{-0.14}^{+0.17}$ \\
UGCA 145 U3    & $18.7_{-2.0}^{+2.3}$ & $0.35_{-0.03}^{+0.08}$ & $3.25_{-0.37}^{+0.38}$ \\
\hline\hline
\end{tabular}\label{table:AGN}
\end{center}
Different sources are fitted with different models. Only the parameters of the power law component of each source are listed here. See text for discussions on the spectral models and the parameters of other components. $F_{\rm X,power}$ is the absorption-corrected 0.3-8~keV flux.\\
\end{table}

Sources U1, U2, and U3 close to UGCA~145 all have absorption column densities exceeding the Galactic foreground value, indicating significant intrinsic absorptions. X-ray spectra of source U1 and U2 can be fitted with a single power law, but the spectra of U3 are very complicated, including two thermal plasma components (with temperature of $1.05_{-0.03}^{+0.02}\rm~keV$ and $0.26\pm0.02\rm~keV$, respectively) and one possible gaussian from Fe~K line emission. Source U1 does not have any significant optical or near-IR counterparts, while U2 and U3 both have point-like optical and near-IR counterparts. The J, H, K band magnitudes of U2 (U3) are 14.0, 13.5, 13.1 (12.0, 11.4, 11.1). The possibly extended emission around U2 and U3 and the large contribution from thermal emission in the spectra of U3 indicate that these two sources are likely members of a background group or cluster of galaxies, although we cannot rule out the possibility that they are MW sources with distance $\lesssim1\rm~kpc$, assuming their intrinsic K-band luminosity equals to the Sun.

\end{appendices}

\end{document}